\documentclass[preprint,12pt]{elsarticle}

\usepackage{subfig}
\usepackage{color}
\usepackage{xcolor,eso-pic,graphicx}
\usepackage{pifont} 
\usepackage{tikz}
\usepackage{amsmath}
\usepackage{amsfonts}
\usepackage{verbatim}
\usetikzlibrary{arrows,shapes}
\usepackage{empheq}
\usepackage{fancybox}
\usepackage{amssymb}
\usepackage{graphicx}
\usepackage{epsfig}
\usepackage{caption}
\usepackage{float}
\usepackage{alltt}
\usepackage{geometry}
\usepackage{lineno}
\usepackage{placeins}
\usepackage{framed}
\definecolor{mycolor}{rgb}{0.122, 0.435, 0.698}
\def\Alg{Alg.}

\begin{document}

\begin{frontmatter}

\title{A portable platform for accelerated PIC codes and its application to GPUs using OpenACC}

\author[1]{F. Hariri\corref{author}}
\author[1]{T.M. Tran}
\author[2]{A. Jocksch}
\author[1]{E. ~Lanti}
\author[3]{J.~Progsch} 
\author[3,4]{P.~Messmer} 
\author[1]{S. Brunner}
\author[2]{C. Gheller}
\author[1]{L. Villard}

\cortext[author] {F. Hariri\\\textit{E-mail address:} farah.hariri@epfl.ch}
\address[1]{Ecole Polytechnique F\'ed\'erale de Lausanne (EPFL), Swiss Plasma Center (SPC), CH-1015 Lausanne, Switzerland}
\address[2]{CSCS, Swiss National Supercomputing Centre, Via Trevano 131, 6900 Lugano, Switzerland}
\address[3]{NVIDIA Switzerland, Technoparkstr 1, 8005 Z\"{u}rich, Switzerland}
\address[4]{ETH Z\"{u}rich, H\"{o}nggerberg, 8093 Z\"{u}rich, Switzerland}

\maketitle

%%****************************************
% Abstract
%%****************************************

\begin{abstract}
We present a portable platform, called \texttt{PIC\_ENGINE}, for accelerating Particle-In-Cell (PIC) codes on heterogeneous many-core architectures such as Graphic Processing Units (GPUs). The aim of this development is efficient simulations on future exascale systems by allowing different parallelization strategies depending on the application problem and the specific architecture. To this end, this platform contains the basic steps of the PIC algorithm and has been designed as a test bed for different algorithmic options and data structures. Among the architectures that this engine can explore, particular attention is given here to systems equipped with GPUs. The study demonstrates that our portable PIC implementation based on the OpenACC programming model can achieve performance closely matching theoretical predictions. Using the Cray XC30 system, Piz Daint, at the Swiss National Supercomputing Centre (CSCS), we show that \texttt{PIC\_ENGINE} running on an NVIDIA Kepler K20X GPU can outperform the one on an Intel Sandybridge 8-core CPU by a factor of $3.4.$ 
\end{abstract}

\begin{keyword}
Plasma Simulations \sep Particle-In-Cell (PIC)\sep GPU \sep OpenACC \sep manycore \sep hybrid architectures
\end{keyword}

\end{frontmatter}

%%****************************************
%% Start line numbering here 
%%****************************************
%\linenumbers

%***********************************************************************
\section{Introduction}
%***********************************************************************
%
Particle-In-Cell (PIC) simulations have been intensively used in many scientific areas, most notably in plasma physics to solve the Vlasov-Maxwell problem or reduced forms such as the gyrokinetic equations. The fundamentals of the PIC method were laid down in the late 1960s and early 1970s and there are a number of textbooks available on this subject~\cite{birdsall2004plasma, hockney1988computer}. Numerical particles are used in the simulation to represent the physical particle distribution function by sampling the phase space in a Monte Carlo sense. Each numerical particle is evolved along the physical particle characteristics using a Lagrangian approach. The charge and current of each particle are deposited on neighboring grid points according to a specific interpolation scheme. After all the charge and current contributions are summed up in this way, the equations for the electromagnetic fields  are solved using a grid-based solver. The calculated electromagnetic fields are then interpolated back to the particles positions, which can then be advanced for another discrete time-step using an ordinary differential equation solver. 
 
Although the PIC method is powerful for handling nonlinear systems, it is demanding in terms of computing power, since a large number of particles is needed to represent the underlying physics. In plasma physics, PIC simulations are an excellent tool to learn about the behavior of magnetized plasmas in tokamaks. However, to simulate plasma turbulence in large tokamaks such as ITER, degrees of freedom on the order of $10^{10}$ particles are required~\cite{villard2013global}. High performance computers offer a powerful tool for gaining an understanding of such complex systems. Along this direction, many strategies to optimize PIC codes have been studied and implemented on parallel architectures~\cite{liewer1989general, bowers2001accelerating, decyk2007upic, tskhakaya2007optimization, stantchev2008fast, decyk2011adaptable}.

As noted in Ref.~\cite{schulthess2015programming}, computers of the twenty-first century have clearly deviated from the von Neumann architectural model, making programming and parallelization in physics more complex. One should keep in mind that architectural diversity and complexity will likely continue to increase. Looking into the near future, this poses a twofold challenge. On the one hand, algorithms and methods need to be designed and improved to make optimal use of new architectures; on the other hand, the portability and good performance of existing codes need to be ensured. In other words, to make use of this type of computing resources, one needs efficient simulation codes to exploit massive levels of parallelism. The subtle issue becomes the portability and good performance of existing codes on complex many and multi-core architectures. 

We have developed a portable platform, called \texttt{PIC\_ENGINE}, for optimizing PIC codes on heterogeneous architectures using NVIDIA GPU accelerators or Intel Xeon Phi processors. This platform serves as a test bed for trying different algorithmic options and parallelization strategies on various hardware platforms. The aim is to provide scientists with performing tools that can be subsequently adapted to specific application codes. With this, it is hoped to achieve efficient simulations on future exascale systems by allowing different parallelization strategies depending on the application problem and the specific architecture. To this end, this platform contains the basic steps of the PIC algorithm in their simplest form and has been designed as a test bed for tuning between different algorithmic choices and data structures. As will be pointed out later in the paper, the way to store data affects performance due to data locality issues. Dealing with machine heterogeneity requires programming models abstract enough to be portable between these architectures, while allowing the compiler to generate efficient code on each platform. A range of such models have been developed in the past, all with different levels of performance portability and compiler maturity. Among the architectures that this engine can explore, we chose to concentrate on systems equipped with GPUs in this work. For current GPU programming, NVIDIA has developed a programming model called CUDA~\cite{cuda}. However, porting of legacy CPU-based applications with CUDA often necessitates explicit compute and data management, thus requiring significant structural changes to existing applications. Therefore, we make the choice to use OpenACC~\cite{openacc}, a directive-based approach allowing to considerably simplify the exploitation of the accelerated processor compared to CUDA. 

In this paper, we demonstrate the feasibility of a portable and efficient PIC implementation based on OpenACC. We provide details of the different parallelization strategies and we show a theoretical understanding of the numerical performance measurements. Using the Cray XC30 system, Piz Daint, at the Swiss National Supercomputing Centre (CSCS), we show that the \texttt{PIC\_ENGINE} code running on a Kepler K20X GPU can outperform this same code running on a Sandybridge 8-core CPU by a factor of $3.4$. In the following section, we further explain our choice of using OpenACC as an alternative to CUDA. In section~\ref{sec:picengine}, the different algorithms used in the test bed and their implementation are detailed. Our expectations from the theoretical performance model are presented in section~\ref{sec:performance model}. Finally, results on a single node are discussed in section~\ref{sec:results}, followed by a discussion in view of physical applications.
%%%%%%%%%%%%%%%%%%%%%%%%%%%%%%%%%%%%%%%%%%%%%%%%%%%%%%%
%%%%%%%%%%%%%%%%%%%%%%%%%%%%%%%%%%%%%%%%%%%%%%%%%%%%%%%
%
%***********************************************************************
\section{OpenACC as an alternative to CUDA}
%***********************************************************************
%
Since the advent of GPUs as general purpose computational architectures, CUDA has become the most widely adopted programming platform for GPU computing, guaranteeing the best possible performance on NVIDIA’s architectures and benefiting from an excellent support from the vendor. Recently a number of PIC algorithms have been successfully implemented on the GPU using CUDA~\cite{decyk1995skeleton, stantchev2008fast, burau2010picongpu, abreu2011pic, meng2013heterogeneous}, efficiently exploiting the accelerator. Such performance gain, however, comes at the expense of adopting a highly complex programming model, which, in general, makes the re-design, re-factoring and subsequent maintenance of the code hard and time consuming. Even worse, the code becomes not portable, since CUDA can run only on NVIDIA GPUs equipped processors. No other architectures are supported. Finally, CUDA is not an open standard, making it difficult to contribute to and influence its development and evolution.

The OpenACC programming model, which has been adopted for the current work, addresses all these issues, promoting an open standard quickly evolving in order to support users’ and, in particular, scientific users’ needs. It proposes a directives based solution similar to OpenMP (toward which it may finally converge). Directives take care of data transfers between CPU and accelerators and manage the work on the accelerator in a fairly simple way, leaving all the inherent complexity to the compiler. The impact on the code’s architecture is often limited, major refactoring being usually necessary only for performance tuning. The same code can run on the CPU and on the accelerator, making its maintainability substantially easier. Furthermore, OpenACC in principle targets any accelerators (not only GPUs), making the code potentially portable on a variety of different architectures. The main drawback of this approach is represented by the performance penalty deriving from the lack of full control on the code performance tuning, and on the strong dependency on the compiler. 

The rest of the section summarizes the main motivations for adopting OpenACC~\cite{openacc} for our specific application as an alternative to CUDA~\cite{cuda}. For further information, the OpenACC standard is available from Ref.~\cite{openacc}.
%
%One of the first parallel PIC codes on shared-memory computers was developed by Horowitz at NERSC~\cite{horowitz1989qn3d}. PIC codes were then developed for distributed-memory computers, a variety of which have already been successfully implemented on GPU platforms in recent years~\cite{decyk1995skeleton, stantchev2008fast, burau2010picongpu, abreu2011pic, meng2013heterogeneous}. In the latter applications, the most computation intensive parts were converted to CUDA, the most widely used technology for developing applications for GPUs. But converting an application to CUDA with the aim of enhancing performance involves significantly more effort than adding directives. This section gives a brief overview of the main reasons why we chose OpenACC~\cite{openacc} as an alternative to CUDA~\cite{cuda}. For further information, the OpenACC standard is available from Ref.~\cite{openacc}. 
%
Launched in 2011, the OpenACC application program interface is a high- level programming model for today's accelerators, allowing large regions of code to be rapidly ported. Concerning supported platforms, up to now the existing OpenACC implementations only support GPUs (from NVIDIA and AMD) and multi-core x$86\_64$ (beta), although support for Intel Xeon Phi is being currently introduced. It is based on using directives for offloading computation and data from the host CPU to accelerators, and it offers portability between various compiler vendors: CRAY and PGI. Also GNU is on the way to integrate it. The biggest virtue of OpenACC that we experienced in our study, is the significant ease of programmability and portability among compilers and devices. Moreover, from a portability perspective, the same code should still run efficiently on a pure CPU-based system. This also allows debugging on the CPU, for which better tools are currently available.

Data transfers between CPU and GPU memory can present a significant obstacle to performance in GPU-accelerated codes. For effective use of the GPU, data transfers between the host processor and the accelerator must ideally be kept to a minimum. Specific OpenACC directives are used to manage these transfers. Other OpenACC directives are used to execute specific regions of the code on the GPU, typically the most compute-intensive ones. In particular, the GPU offers massive amounts of multi-threading (for K20X, there are specifically 2048 active threads per Streaming Multiprocessor and there are 14 Streaming Multiprocessors). These threads are organized into a hierarchy of three levels. At the coarsest level are the thread blocks. At the second level, the threads are grouped into so-called ``warps". Finally, we have the finest level threads. These 3 levels of multi-threading can be controlled by adding clauses (\texttt{gang}, \texttt{worker}, \texttt{vector}) in OpenACC directives.   

Examples of accelerating the charge deposition routine in a PIC code can be found in~\ref{app_B}.\\

The OpenACC programming model allows large regions of code to be rapidly ported with minimal effort. For applications which already have optimized CUDA kernels, we recall that OpenACC supports CUDA, allowing the usage of highly specialized functions available in CUDA only, of course at the expense of portability and maintainability. We will later show in section~\ref{sec:results}, that the result of combining OpenACC and CUDA can provide the best performance.\\

In this paper, our focus is to use OpenACC as the main programming model for accelerating the \texttt{PIC\_ENGINE} on hybrid architectures. Although the need to move data and parallelize loops is intrinsic to accelerated computing platforms and therefore common to both OpenACC and CUDA, the programming easiness and portability favors OpenACC. We performed low-level CUDA optimizations only to the most costly routines. However, for these routines, an OpenACC version is also available, to guarantee portability on non-NVIDIA platforms. In general, in order to ensure portability and maintainability, proprietary solutions, like CUDA, should have a limited usage, in favor of open standards like OpenACC and OpenMP, in view of their possible future convergence.
%%%%%%%%%%%%%%%%%%%%%%%%%%%%%%%%%%%%%%%%%%%%%%%%%
%%%%%%%%%%%%%%%%%%%%%%%%%%%%%%%%%%%%%%%%%%%%%%%%%
%
%***********************************************************************
\section{The PIC\_ENGINE test bed}\label{sec:picengine}
%***********************************************************************
%
In plasma physics applications, the PIC algorithm is used to describe the time evolution of the distribution functions of different particle species (ions, electrons, impurities) in their self-consistent electromagnetic fields. The \texttt{PIC\_ENGINE} is a three-dimensional PIC code written in Fortran. It is developed for hybrid architectures to provide a test bed for evaluating performance of different algorithms and methods. This code has been deliberately kept to a minimum, and it embeds the main pieces underlying the PIC method. In a typical simulation of plasma dynamics, the system is represented by a set of numerical particles. For each particle one needs to store the following attributes: position and velocity $(\mathbf{x}, \mathbf{v})$ that are initialized at the start of the simulation. Particles interact with each other through the self-consistent electromagnetic field. The following four consecutive subroutines are executed at each time step for all particles in each cell on the mesh:
\begin{itemize}
\item[i.] \texttt{setrho()}: $(\mathbf{x}) \longrightarrow \rho$ 
\item[ii.] \texttt{field\_solver()}: $- \Delta \phi = \rho$ 
\item[iii.] \texttt{accel()}: ${\mathbf{v}^{t+1}} = {\mathbf{v}^{t}} + \Delta t \, \mathbf{E}^t$ 
\item[iv.] {\texttt{push()}}: ${\mathbf{x}^{t+1}} = {\mathbf{x}^{t}} + \Delta t \, \mathbf{v}^{t+1}$
\end{itemize}

In step i. the deposition of the charge density $\rho$ (called gather operation) of each particle on the grid is called \texttt{setrho()} using the so-called particle weighting~\cite{birdsall2004plasma}. Prior to the calculation of the forces on each particle from the electric potential, we solve Poisson's equation for computing the field potential $\phi$ in \texttt{field\_solver()}. This information is then used for moving the particles in time according to the equations of motion. The particles velocities are thus updated accounting for the $\mathbf{E}$ field in \texttt{accel()}. This involves an interpolation (called scatter operation) to obtain the fields at the particle position from the fields on the nearest grid points. Going from grid-based data to particle-based data, required for estimating $E$ at the particle positions, actually explains the separation of \texttt{push()} and \texttt{accel()} into two distinct steps. Finally, in the fourth step of the algorithm, called \texttt{push()}, the particle positions are updated.\\ 

\begin{figure}[h!]
	\centering
       \includegraphics[width=0.4\linewidth,keepaspectratio,clip]{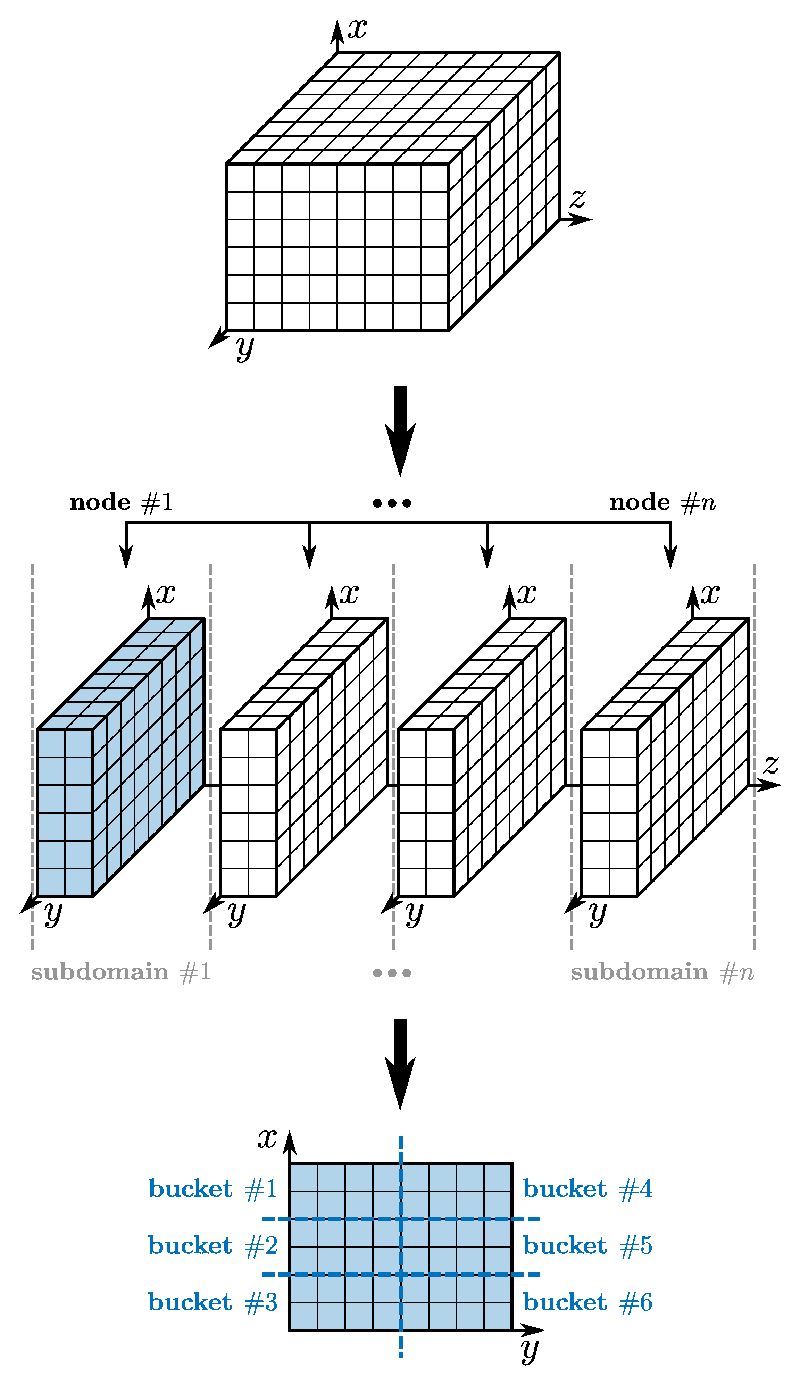}
       \caption{In this figure we show the domain decomposition of the \texttt{PIC\_ENGINE} 3D grid space $(\mathbf{x}, \mathbf{y}, \mathbf{z})$ on $n$ nodes. Only one slice (in blue) is considered in the present study for benchmarks on a single node. In an $(\mathbf{x},\mathbf{y})$ plane, particle data are sorted into buckets, each containing a certain number of grid cells.}
       	 \label{fig:domain_decomp}
 \end{figure}

For the sake of simplicity, the version considered for this study uses the full-f PIC method applied to the Vlasov electrostatic problem in 6D phase space using Cartesian coordinates. Boundary conditions are periodic. Magnetic fields are neglected. Step ii involving \texttt{field\_solver} is not considered. The electric field at the particle's position $\mathbf{E(x)}$ is assumed to be constant. Particles are pseudo-randomly loaded uniformly between the system bounds and linear B-splines are used for particle-to/from-grid operations. We consider a uniform 3D cartesian grid in real space $(\mathbf{x}, \mathbf{y}, \mathbf{z})$, shown in Fig.~\ref{fig:domain_decomp}. Domain decomposition is applied in the $\mathbf{z}$ direction where each domain is assigned to a different compute node with inter-node communication using MPI. In this paper, we focus on parallelization strategies on a single node using OpenACC.
It is important to note that the \texttt{PIC\_ENGINE} is very simple with respect to realistic PIC codes having significantly more arithmetic operations. For example, in plasma physics simulations, gyrokinetic PIC application codes such as ORB5~\cite{tran1998global, jolliet2007global} and GTC~\cite{lin1998turbulent,ethier2005gyrokinetic,ethier2008large} have many more arithmetic operations in the \texttt{accel} and \texttt{push} routines than \texttt{PIC\_ENGINE}. Also \texttt{setrho} involves higher order B-splines and therefore, together with \texttt{accel} and \texttt{push}, dominates the timings. This engine is intended to be used as the core of actual PIC application codes where its components will be respectively called depending on the physics problem and the given architecture.
%
%%-----------------------------------------------------------------
\subsection{Data structures}
%%-----------------------------------------------------------------
%
PIC codes have two main data structures, particles and fields, that need to communicate with one another. Particles are a priori randomly distributed in space. As a consequence, particle data is totally unstructured leading to pseudo-random memory access for the grid-to-particle and particle-to-grid operations. The challenge to address at this level concerns organizing particle data structures so that data locality is improved both in the grid-to-particle interpolation step, i.e. getting the electric field at the particle position in \texttt{accel}, and in the particle-to-grid operations in the charge deposition step \texttt{setrho}. For each particle, we assign six attributes: three coordinates for the position $(x,y, z)$ and three velocity components $(v_x,v_y,v_z)$. The grid is uniform in real space $(x_i, y_j, z_k)$.

For an identical numerical algorithm the particle data can be stored in multiple ways. The first option is to store particles into multidimensional bins defined according to their position with respect to the grid and to keep particles in each bin stored together in memory. This procedure, called \emph{binning} has been used in~\cite{stantchev2008fast, maduri_etal:2011, decyk_singh:2014}. It requires reordering of particles into bins after every particle \texttt{push()}. Each bin can then be processed independently in parallel. The particle data is represented in this case by the particle array \texttt{part\_att(max\_numpart,natt,tot\_bins)}, where \texttt{max\_numpart} is the maximum number of particles per bin, \texttt{tot\_bins} is the total number of bins and \texttt{natt} is the number of attributes here equal to $6$ ($3$ positions and $3$ velocities). An additional array \texttt{npbin(tot\_bins)} is used to specify the number of particles in each bin.
Binning is advantageous as particles in specific bins can be accessed directly without the need of offset variables.
And if the number of particles in a bin changes, then there's no need for rearrangement of other particles in memory. 

With this approach, memory usage is not optimized. Memory has to be over-allocated and gaps between the particles of the different bins occur.
Thus the alternative option is to store particle data contiguously in memory and declare it as follows: \texttt{part\_att(num\_part, natt)}.
Where \texttt{num\_part} is the number of particles and \texttt{natt} refers to the number of attributes.
In this second data layout, the particles are stored contiguously and have a priori an arbitrary order in the array.
The contiguous data structure has the advantages that it is optimal with respect to memory requirements, no gaps exist in memory.
In order to get better performance, one might improve the data locality in the contiguous structure by sorting the array of particles into buckets which are defined with respect to the grid (see next subsection).

\begin{figure}[h!]
	\centering
       \includegraphics[width=0.8\linewidth,keepaspectratio,clip]{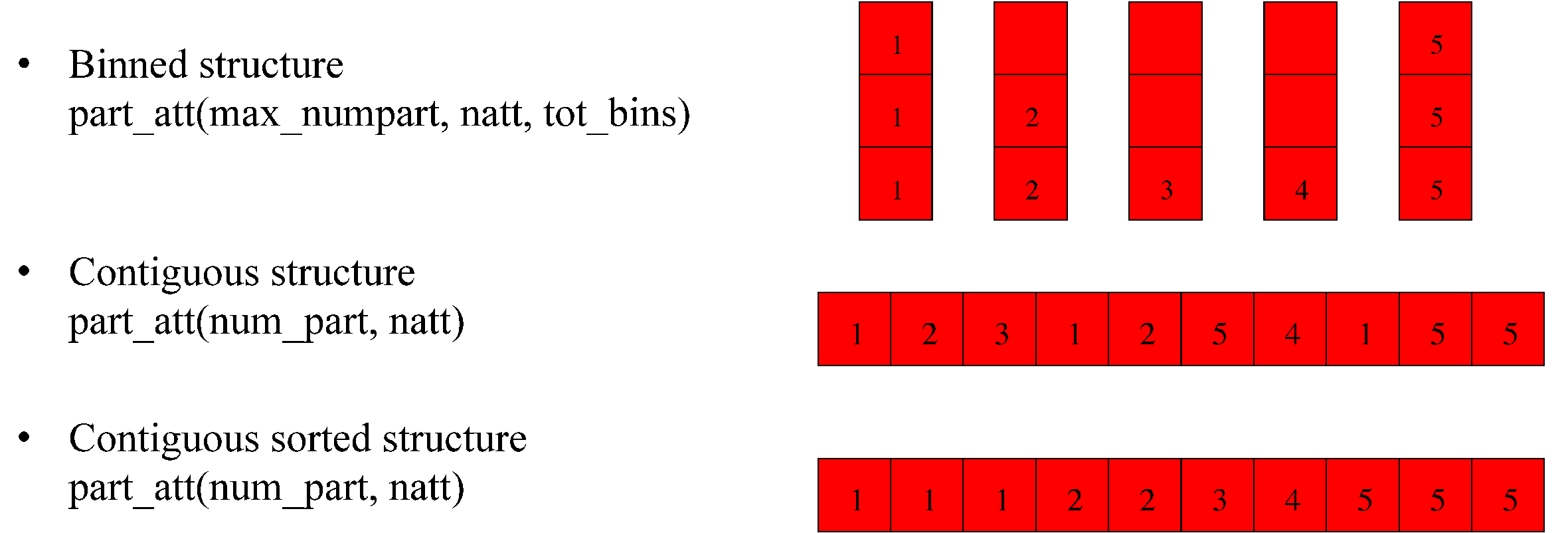}
       \caption{different data structures in the code: binned, contiguous, and contiguous sorted. The numbers correspond to the particle key index.}
       	 \label{fig:datastruct}
 \end{figure}

Moreover, for the above two data layouts, the particle positions and velocities stored in the \texttt{part\_att} array can be organized in two different ways. The first way is to store particles attributes in rows and every column points to a different particle, like for example \texttt{part\_att(1:natt,1:num\_part)}. This is the so-called array of structures (AOS). The second way is to assign each row to a particle and columns will contain each particle's attributes, like \texttt{part\_att(1:num\_part,1:natt)}. This is the so-called structure of arrays (SOA). In the \texttt{PIC\_ENGINE}, a Fortran memory layout where particle data is stored as a Structure of Arrays (SOA) showed higher performance on both GPU and CPU. We thus consider it for our benchmarks. The two data layouts that we consider for storing particle data \texttt{part\_att} in arrays are illustrated in Fig.~\ref{fig:datastruct} where the particle's key index written in each cell corresponds to the target bucket assigned to that particle. 
%
%%-----------------------------------------------------------------
\subsection{Sorting}\label{subsec:sorting}
%%-----------------------------------------------------------------

The concurrency of particle-to-grid operations occurs when particles deposit their charge on the same grid point. This constitutes a key challenge for parallelizing the PIC algorithm. While the fields are fixed to the mesh lattice and can be stored on a local subset of the memory, the field update depends on the charge density updated from the particle positions. Sorting the particles according to their position in real space is necessary for increasing data locality, therefore improving performance.
We sort the particles for the different PIC implementations we consider.
The particles are sorted with respect to a partition of the grid into subdomains that, in the remainder of this paper, will be called "buckets". A bucket can consist of a single grid cell or can contain multiple cells. Depending on how particles are deposited to the grid, we consider different granularity of the sorting with different bucket sizes.
In a realistic implementation of a global plasma domain, particles cross cell boundaries. Therefore, the field data of grid cells along the border between buckets must be transferred at each time step and particle crossings require data sorting at every time step to retain data locality (associating the particles to the bucket they are in). 
In the context of the PIC method on GPUs, various sorting algorithms have been employed.
Decyk and Singh~\cite{decyk_singh:2014} applied particle sorting to a two-dimensional system where it was a constraint that particles can move at most to the nearest neighbouring buckets every time step.
The algorithm of Stantchev et al.~\cite{stantchev_dorland_gumerov:2008} also only exchanges particles between nearest neighbouring buckets. A recursive sorting approach is then suggested to account for fast particles.
Mertmann et al.~\cite{mertmann_etal:2011} implemented a particle reodering method with a bucket sort algorithm.
Joseph et al.~\cite{joseph_etal:2011} investigated a particle in cell method based on a triangular mesh where a bucket sort algorithm was applied to the particles.
Rozen et al.~\cite{rozen2008gpu} implemented a bucket sort algorithm using linked lists.
 
For our application, we use two different implementations of the bucket sort algorithm where particles are sorted according to their keys (which represent their target bins).
Our first option is the standard bucket sort algorithm:
For the binned data structure a second temporary particle data array of the same size and structure as the primary particle data array is used together with a field of counters which represent the number of particles in each bucket.
All counters are set to zero. Afterwards all particles of the primary particle data array are moved in their correct bins in the temporary particle data array.
Counters give the positions in the bins which are updated \verb+atomically+ for every particle moved. That is, \texttt{atomic} operations are used to resolve collisions of threads trying to access the same address in memory.
Thus the particles are sorted.
Since the standard bucket sort is an out-of-place sorting algorithm (that is results are at a different location in memory than the original data), and since we want the particle data in the primary array, the temporary array is copied back to the primary particles array.

For the contiguous data layout we apply the bucket sort implementation of Sintorn and Assarsson presented in~\cite{sintorn_assarsson:2008}. All particles of the array are divided into groups assigned to different threads. For every particle, a counter of the target bin is incremented using atomic directives in order to determine the bin sizes. At the same time the value of the counter is stored in an array at the particle's position. The particles are copied to the target array, where the offset for every particle is the sum of the corresponding counter and the stored value. The target array then contains the sorted particles. For this out-of-place method a second array of identical structure is required. Since the particles are required in the primary array, they need to be copied back to the source array. For our application this algorithm is faster than the radix sort algorithm provided by the Thrust and CUDPP libraries~\cite{merrill_grimshaw:2011}.

\begin{figure}[h!]
   \centering
   \includegraphics[width=0.8\linewidth,keepaspectratio,clip]{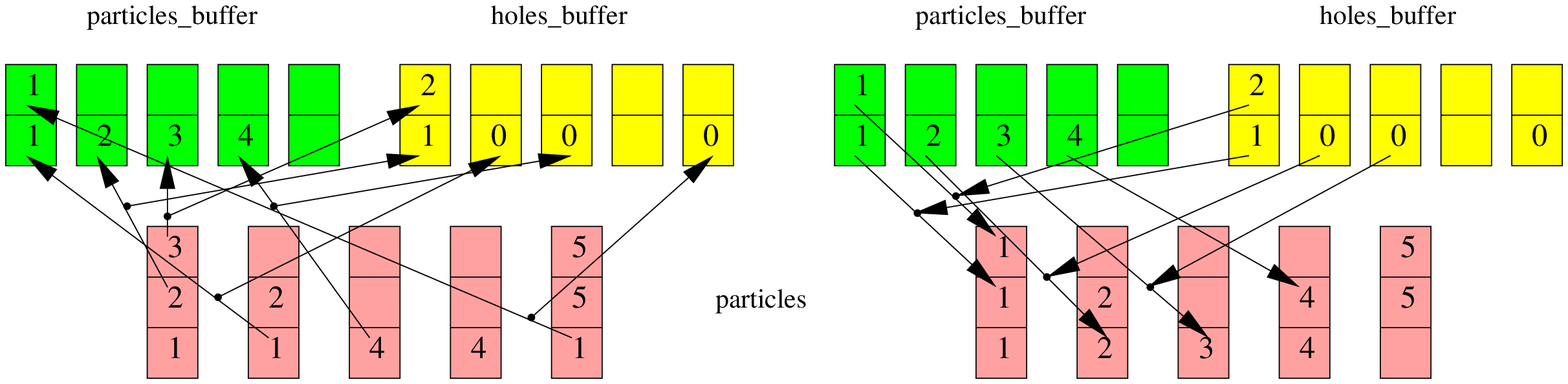}
   \caption{Bucketsort for pre-ordered particles, in case of binned datastructure. Numbers are particle keys execept for holes buffer where they are offsets}
   \label{fig:bucketsort_binned}
\end{figure}

Depending on the size of the buckets, the time step, the particles velocity and the geometry, the fraction of particles moving between buckets within one time step varies.
If we are in a regime where only a small portion of particles moves between buckets, we apply a bucket sort algorithm which exploits that the particles are pre-ordered. This is explained in detail in Ref.~ \cite{jocksch2015bucket}.
The method is more efficient than a standard bucket sort because only data of particles which change their bucket are moved. For the rest of this paper, this sorting algorithm is considered. It is shortly summarized here:

For the binned data structure (Fig.~\ref{fig:bucketsort_binned}, Algorithm 1) a particles buffer (\texttt{particles\_buffer}) of sufficient size (we chose the binsize of the particles array) is allocated for every bin.
\begin{figure}[h!]
   \centering
   \includegraphics[width=0.8\linewidth,keepaspectratio,clip]{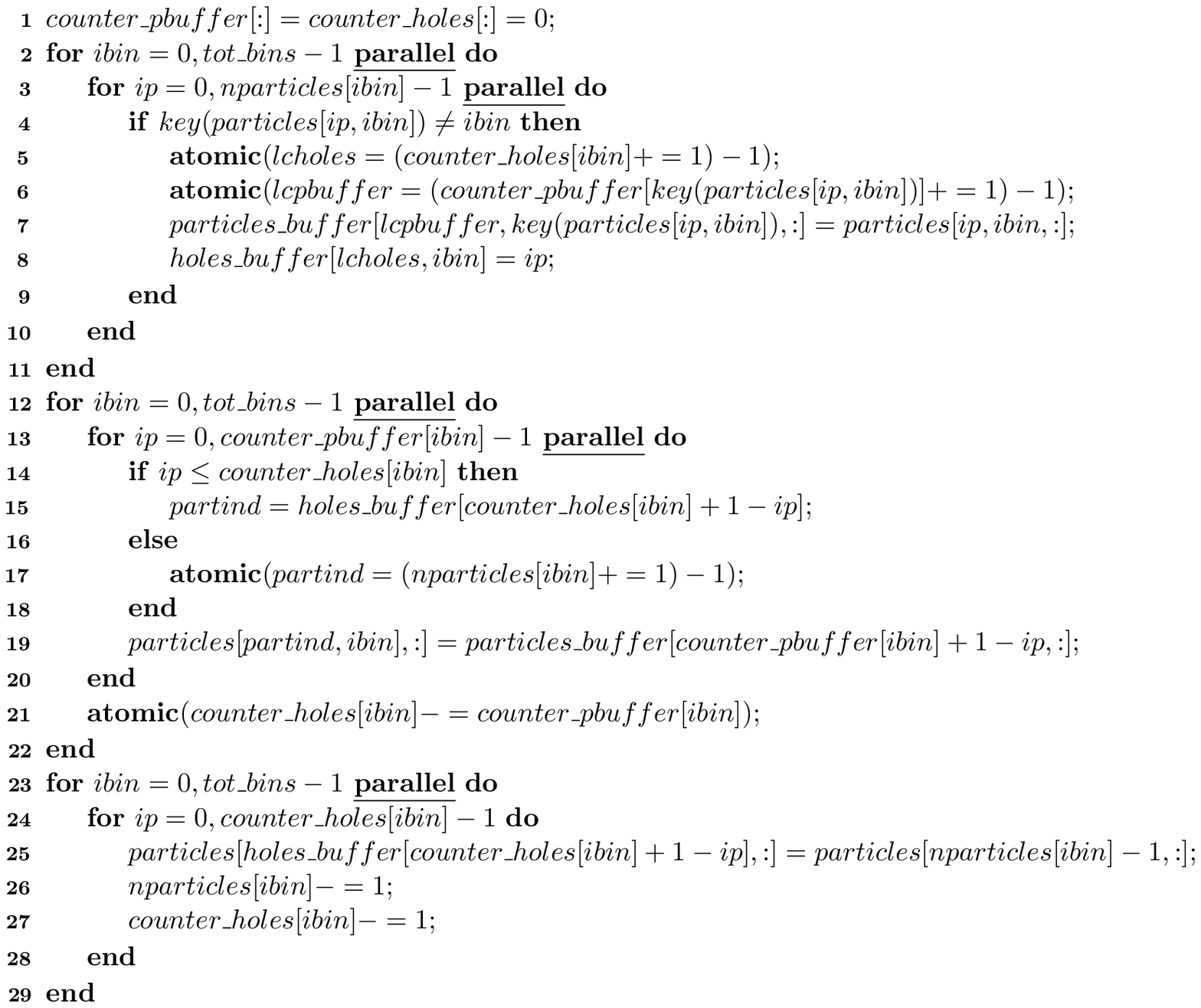}
   \caption*{\Alg{}~1: Bucketsort for pre-ordered particles, binned datastructure}
   \label{alg:bucketsort_binned}
\end{figure}
Furthermore for every bin a buffer for the holes (\texttt{holes\_buffer}) is allocated (we chose the same size as for the particles buffer) where the indices of the particles having to be moved are stored.
For the indices in the particles buffer and the holes buffer, counters are allocated and set to zero.
All particles which are not in their target bin are moved into their target buffer and the hole is marked in the holes buffer of the source bin (Fig.~\ref{fig:bucketsort_binned} left).
The counters for the indices in the particles buffer and the holes buffer are incremented  \verb+atomically+.
Then the particles in the particles buffer are moved to their target bins (Fig.~\ref{fig:bucketsort_binned} right) at the location of the holes in holes buffer (Fig.~\ref{fig:bucketsort_binned} the first three bins) or, if no holes are marked, to the end of the particles buffer (Fig.~\ref{fig:bucketsort_binned} the fourth bin).
Afterwards the remaining holes are compacted (in Fig.~\ref{fig:bucketsort_binned} the fifth bin needs to be compacted) and the particles are sorted.

\begin{figure}[h!]
   \centering
   \includegraphics[width=0.8\linewidth,keepaspectratio,clip]{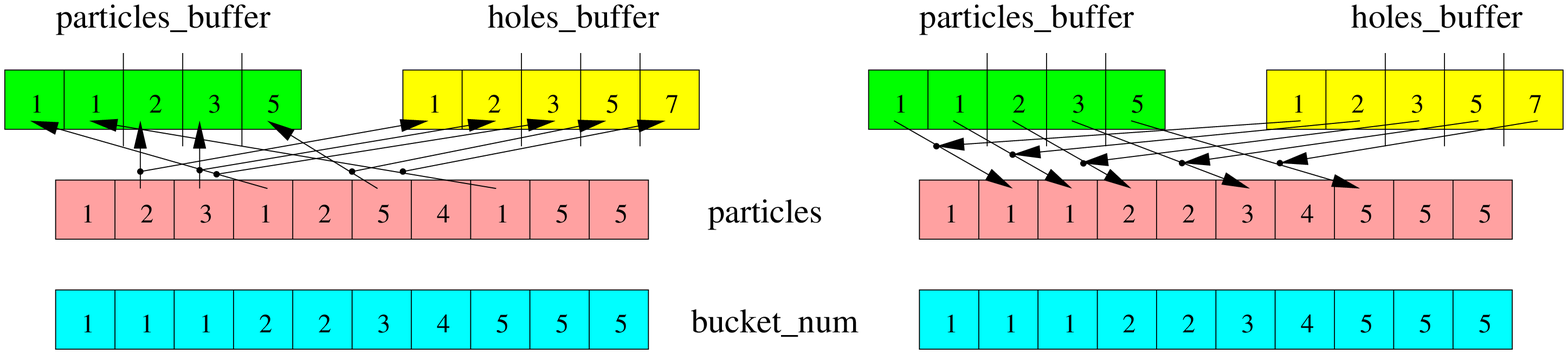}
   \caption{Bucketsort for pre-ordered particles, contiguous datastructure, numbers are particle keys except for holes buffer where they are offsets}
   \label{fig:bucketsort_con}
\end{figure}

For the contiguous memory layout (Fig.~\ref{fig:bucketsort_con}, Algorithm 2) all particles are counted with respect to their key, a histogram is computed.
\begin{figure}[h!]
   \centering
   \includegraphics[width=0.8\linewidth,keepaspectratio,clip]{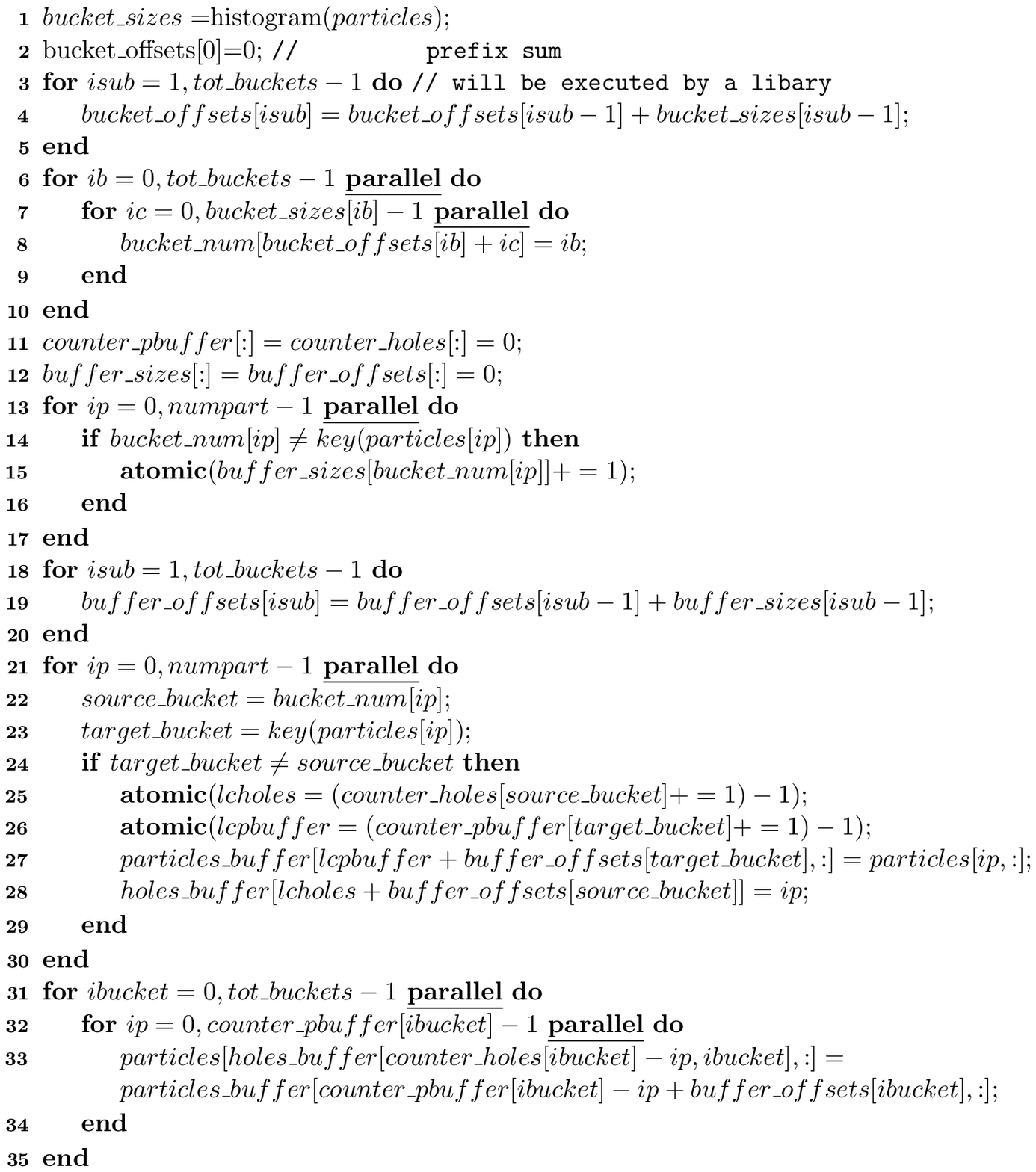}
   \caption*{\Alg{}~2: Bucketsort for pre-ordered particles, contiguous datastructure}
   \label{alg:bucketsort_con}
\end{figure}
With the histogram data a reference array \texttt{bucket\_num} with the sorted keys is established.
The number of particles leaving each bucket is counted (second histogram) in order to determine the size of the buffers and the offsets for the buckets in them.
Two buffers are used: one for the particles (\texttt{particles\_buffer}) and one for the holes left in the particle array if particles are taken out (\texttt{holes\_buffer}).
In addition, variables for the number of entries in the buffers are used.
All particles which are not in their bucket are moved to the particles buffer and the holes are registered in the holes buffer, the variables for the number of entries are incremented with \verb+atomic+ operations (Fig.~\ref{fig:bucketsort_con} left).
After this, the particles are put into their target buckets according to the entries in holes buffer (Fig.~\ref{fig:bucketsort_con} right).
Thus the particles are sorted.

%%-----------------------------------------------------------------
\subsection{Parallel charge assignment}\label{sec:setrhoroutines}
%%-----------------------------------------------------------------
%
Calculating the charge density $\rho$ is carried out by depositing charge, assuming a linear particle shape, from the set of scattered particle positions onto the fixed spatial grid. This operation constitutes the first step of the PIC code, and is referred to as the charge deposition. It is also the most critical and computationally challenging step because race condition can happen when two threads attempt to write to the same memory location simultaneously. This requires either synchronization or atomicity of the operations to resolve it. Otherwise the result can be arbitrarily wrong. In this section, we will present different approaches to parallelize the deposition step. We distinguish between two types of algorithms: collision-free algorithms and collision-resolving algorithms. Note that for all options one can have either binned or contiguous data structures. 
%%%%
\subsubsection{collision-free: threads on splines \label{subsub:threadsgrid}}
\begin{figure}[h!]
  \centering
  \includegraphics[width=0.75\linewidth,keepaspectratio,clip]{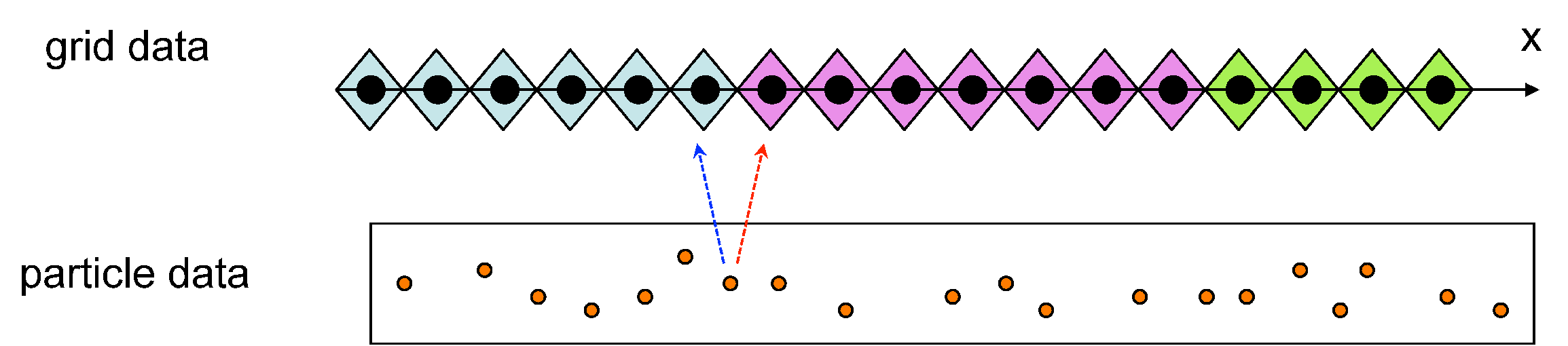}
  \caption{Parallelize the charge assignment routine with threads on the grid}
  \label{parallel_approach_2}
\end{figure}

\begin{figure}[h!]
  \centering
  \includegraphics[width=0.75\linewidth,keepaspectratio,clip]{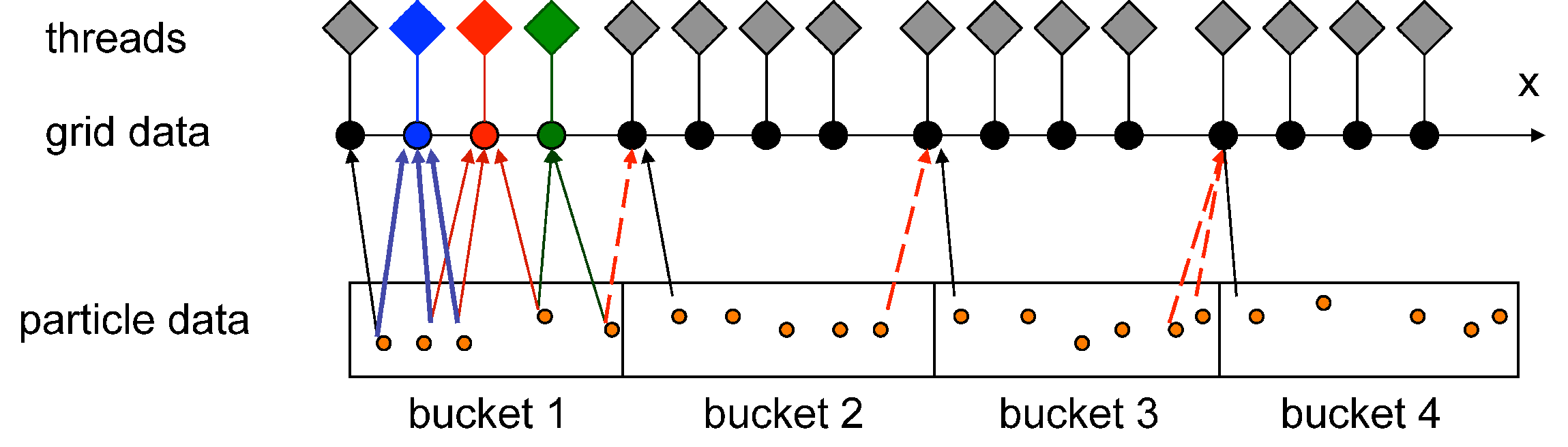}
  \caption{Parallelize the charge assignment routine with  threads on the grid and bucket-sorted particles}
  \label{parallel_approach_4}
\end{figure}

%%%%
The first approach consists of associating the threads with the grid data. We refer to this routine as \texttt{setrho\_splines()}. In this case, different threads may read the same particle data, but do not need to update the particle data. With this procedure, we avoid race condition and synchronization is no more necessary. A $1$D illustration is given in Fig.~\ref{parallel_approach_2}. 
This approach can be costly if each thread loops over all particles to read data. One remedy is to have data sorted in buckets, as shown in Fig.~\ref{parallel_approach_4}. By counting the loop iterations, the cost of this routine is found to be proportional to
  \begin{equation*}
    \label{eq:splinescost}
  (N_x+d)(N_y+d)(N_z+d)\frac{N_p}{N_{buckets}},
  \end{equation*}
where $N_p$ is the total number of particles. $N_x, N_y$, and $N_z$ are respectively the number of grid intervals in $\mathbf{x}, \mathbf{y}$ and $\mathbf{z}$. $N_{buckets}$ is the total number of buckets, and $d$ is the order of the interpolation ($d=1$ in our case). So the \emph{minimum} cost is obtained by \emph{maximizing} the number of buckets for fixed grid size and number of particles, i.e. for buckets containing just one grid cell.
\FloatBarrier
This algorithm can be applied with either binned or contiguous data structures.
%%%%
\subsubsection{collision-resolving: threads on buckets\label{subsub:threadsbins}}
In this approach, one thread is assigned per bucket. The particles should  be fully sorted, i.e: buckets should contain one grid cell only. The main particle loop is now replaced by an outer loop over the buckets, and an inner loop over the particles within each bucket. The outer loop is partitioned across the \emph{threadblocks} while the inner loop is serial. Since different threads of the outer loop write to the same grid points, parallelization on the outer loop on buckets requires the \texttt{atomic} directive to resolve collisions. However, for higher order interpolation, this approach would utilize a large amount of shared memory (knowing that the shared memory on a K20 GPU is 48KB per thread-block). 
%%%%
\subsubsection{collision-free: threads on buckets with data replication}
\begin{figure}[h!]
  \centering
  \includegraphics[width=0.75\linewidth,keepaspectratio,clip]{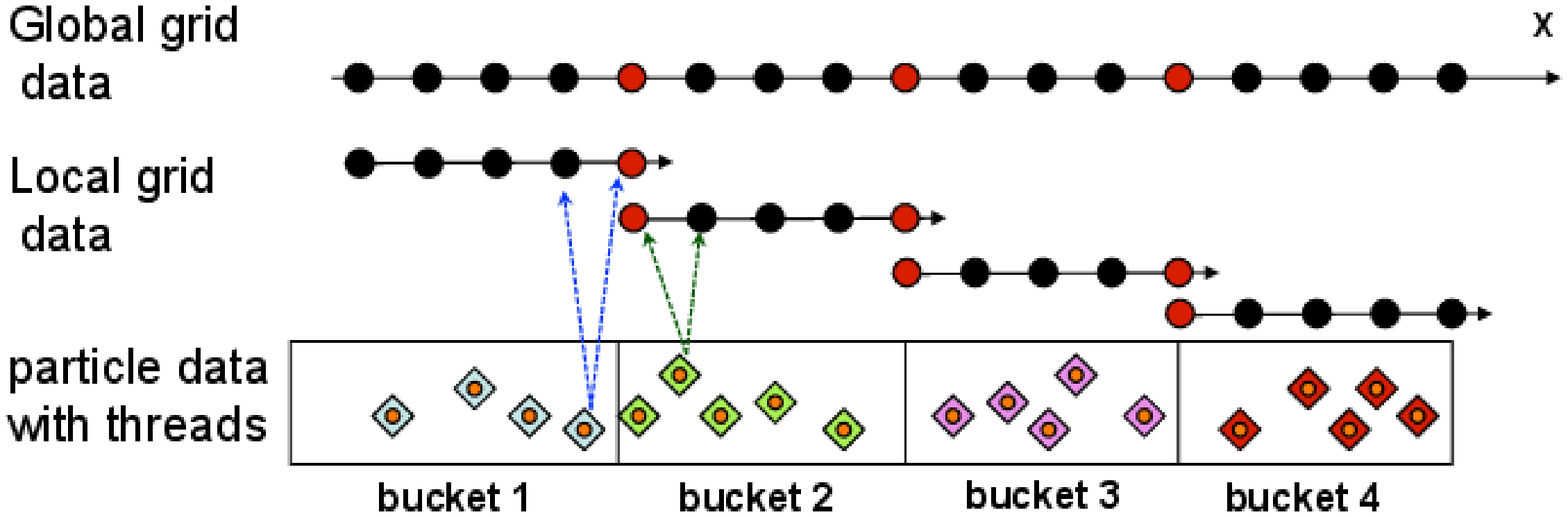}
  \caption{Parallelize with threads on buckets of particles}
  \label{parallel_approach_3}
\end{figure}

The approach is similar to the one in sec.~\ref{subsub:threadsbins} but \texttt{atomics} are not used. Instead, as suggested in~\cite{decyk_singh:2014}, one solution is to replicate the grid data from global to local data and sort particles in buckets (according to their position on the grid). An example is illustrated in Fig.~\ref{parallel_approach_3}. The parallel approach consists of associating threads with buckets of particles. This ensures that each thread does the charge assignment on its local grid thus preventing race condition. Guard-cells defining the border of buckets (red points in Fig.~\ref{parallel_approach_3}) provide temporary memory for data exchange. They are added separately to the global grid data. The outer loop on buckets is collision-free by constructing the charge on \emph{local spatial subdomains} which include guard cells, shared with their right neighbor subdomain, in each of the 3 directions. An additional procedure has to be performed to assemble the global charge deposition. This requires two steps in the algorithm. In the first step, reduction can be applied to local temporary variables (red points in Fig.~\ref{parallel_approach_3}) without atomics in global memory. In the second step, threads are assigned to the grid and collect the data from the temporary variables in global memory. 
%%%
\subsubsection{collision-resolving: threads on particles\label{subsub:threadsparticles}}
\begin{figure}[h!]
  \centering
  \includegraphics[width=0.75\linewidth,keepaspectratio,clip]{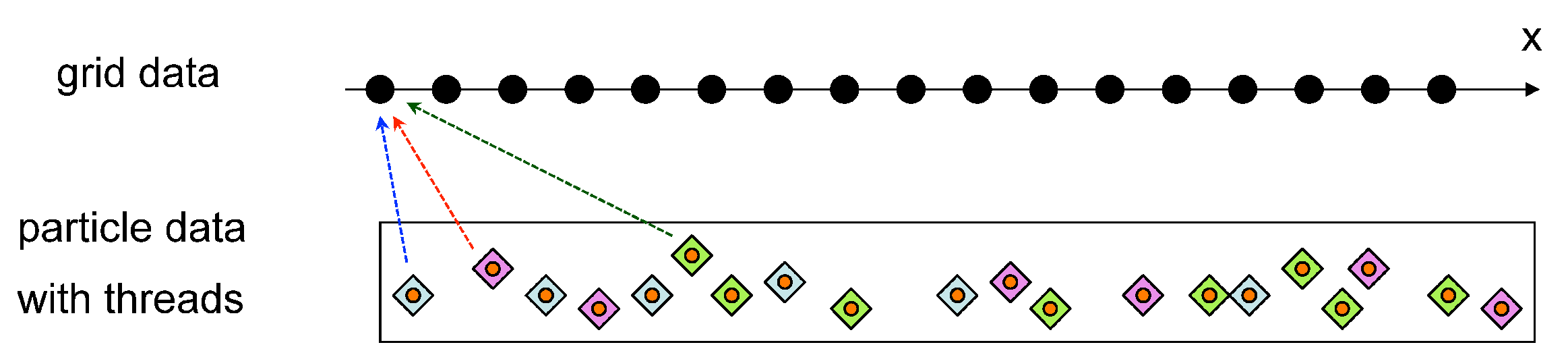}
  \caption{Parallelize with threads on particles}
  \label{parallel_approach_1}
\end{figure}

An illustration of a $1$D case is given in Fig.~\ref{parallel_approach_1} where threads associated with particle data are represented by diamonds of different colors. In this case, different threads can update the same grid data, thus leading to \texttt{race condition}. Synchronization is therefore needed to resolve the conflict. On the NVIDIA GPUs, this concurrency of particle-to-grid operations can be avoided by using \texttt{atomic} operations. One thread is assigned to each particle which uses atomic add to deposit the charge to the grid. An optimization of this algorithm is possible in CUDA by using shared memory as an intermediate step to do the reduction (see example in~\ref{app_B}). In this case, sorting is required and the reduction is done with two steps: first with CUDA to shared memory, then from shared memory to global memory where the grid is stored. 
An alternative to this collision-resolving algorithm avoids direct reduction to the grid.  The reduction from every particle can be done on a temporary thread-private variable without atomics. The result of the temporary variables is put only once with atomic adds to the grid. 
%
%***********************************************************************
\section{Performance model}\label{sec:performance model}
%***********************************************************************
%
This paper examines the particle-in-cell algorithm on a single node of the CRAY XC30, Piz Daint supercomputer, to see how well it performs. Piz Daint consists of 5272 compute nodes each with one Intel$^{\circledR}$ Xeon$^{\circledR}$ E5-2670 @ 2.60GHz CPU (8 cores, 16 virtual cores with hyperthreading enabled, 32GB RAM) and one NVIDIA$^{\circledR}$ Tesla$^{\circledR}$ K20X GPU. 

We present results of a study on modeling the performance of the \texttt{PIC\_ENGINE} components on Piz Daint. We model computation time by counting coalesced memory accesses as the size of the accessed data type, and non-coalesced accesses as 32Bytes~\cite{nvidiadocs}. The GPU executes small groups of threads called warps. The term coalesced refers to adjacent threads (within a wrap) accessing adjacent locations in memory. These are in turn, summed up, and divided by global memory bandwidth. The following calculations assume $10^6$ particles and $512\times256$ grid cells, knowing that the achievable practical global memory bandwidth is about 180 GB/s for a K20X with ECC enabled (Error-correcting code memory which reduces the available usable bandwidth). Recall that the size of a double is 8 bytes compared to 4 bytes for a single or integer precision. We will present our model estimates separately for each routine in double and single precision.
%
%-----------------------------------------------------------------
\subsection{push}
%-----------------------------------------------------------------
%
In this routine we count 6 coalesced loads $(x, y, z, v_x, v_y, v_z)$ and 3 coalesced writes $(x, y, z)$ per particle. The estimated computation time for double precision is $t_{double}= 9\times8 B/(180 GB/s) = 0.4 ns$ and the estimated time for single precision is $t_{single}= 9\times 4 B/(180 GB/s) = 0.2 ns$.
%
%-----------------------------------------------------------------
\subsection{accel}
%-----------------------------------------------------------------
%
In this case we have scattered reads through the GPUs read-only data cache, so the actual global memory transactions depend on the fraction of cache misses (miss rates), which might not be easy to calculate from “first principles”. Assuming working sets are small enough or we have good locality the lower bound on miss rate is: Bytes/Bytes accessed. Or, in other words, the assumption is that each element is loaded exactly once even if used multiple times (this assumes that cache bandwidth is “high enough” not to affect the timings). Here we count 6 coalesced loads $(x, y, z, v_x, v_y, v_z)$, 3 coalesced writes $(v_x, v_y, v_z)$, and 12 scattered reads through read-only data cache, from three grids ($x,y,$ and $z$ components of the electric field) of $512 \times 256$ cells each. We consider $N=10^6$ particles. For perfect cache efficiency, the estimated best case computation time for double precision is $t_{double}= (9 + 3\times 512 \times 256/10^6)\times8 B/(180 GB/s) = 0.417 ns$ and the worst case, with 100\% cache miss rate, computation time is $t_{double}=(9\times8 B + 12\times32 B) /(180 GB/s) = 2.53 ns$. For single precision, the estimated best case computation time is $t_{single}= (9 + 3 \times 512\times256/10^6)\times 4 B/(180 GB/s) = 0.2087 ns$ and the worst case gives $t_{single}=(9\times 4 B + 12 \times 32 B) /(180 GB/s) = 2.3333 ns$. Thanks to the sorting, we shal see that we are almost in the best case.
%
%-----------------------------------------------------------------
\subsection{setrho}
%-----------------------------------------------------------------
%
In the following calculations, the charge deposition algorithm with threads on particles, \texttt{setrho\_3.3.4}, is considered. For the scattered writes that appear in this step we need to use atomic operations since two particles can cause concurrent writes to the same memory location. Since the GPU provides hardware atomics for single precision, we ignore the additional overhead and model the writes as regular scattered (i.e. non-coalesced) writes. In the kernel we count 3 coalesced loads $(x, y, z)$ and 8 scattered atomic writes. Similar to the scattered loads from before we count non-coalesced writes as 32 bytes transactions in terms of consumed bandwidth. The estimated computation time for double precision is $t_{double}= (3\times8 B + 8\times32 B) /(180 GB/s) = 1.56 ns$ and the one for single precision is $t_{single}= (3\times4 B + 8\times32 B) /(180 GB/s) = 1.49 ns$.

In the case of \texttt{accel} we also considered the impact of locality and the read-only data cache under the assumption that the cache bandwidth is sufficiently high not to appear in the calculation. The same locality assumption also holds in this case but the assumption for the cache does not since we can not use the read-only data cache for write operations. We might still see a benefit from data being cached in the GPUs L2 cache but the exact impact of that is hard to estimate. One should also note that the used K20X GPU does not support native double precision atomics which have to be implemented in terms of compare and swap operations instead. This estimate may therefore underestimate the cost for double precision. 

%Atomics are executed in L2 cache. L2 bandwidth is about 3 times practical GMEM bandwidth and we can not fully utilize it due to scattered writes only hitting parts of a sector (32Bytes) at a time. The “high enough bandwidth” assumption from the TEX case does not hold here. We count 3 coalesced loads $(x, y, z)$ through L2 and 8 scattered atomic writes to L2. Essentially, the price with atomics is paid per cacheline touched, and the cacheline size is 32 bytes. 8 locations are touched per particle. Knowing that the L2 bandwidth is roughly 562 GB/s, the estimated computation time for double precision is $t_{double}= (3\times8 B + 8\times32 B) /(562 GB/s) = 0.4982 ns$ and the one for single precision is $t_{single}= (3\times8 B + 8\times32 B) /(562 GB/s) = 0.4769 ns$.

%
%-----------------------------------------------------------------
\subsection{psort}
%-----------------------------------------------------------------
%
The cost of the sorting scales with the amount of particles. The practical cost of sorting can be broken down into the cost for reading a particle's data plus the cost of storing if the particle is misplaced. Since the cost of misplaced particles involves expensive non-coalesced writes, it has significant impact on performance even if the misplaced particles are few. Additionally there are histogram and prefixsum operations in the algorithm. But since these are hard to model, they are not incorporated in the model. We define $N$ to be the number of particles and $M$ the number of misplaced particles. In addition to the 8 B double transactions there are 4 B integer transactions and non coalesced loads of both types that “cost” 32 B. The total transactions per particle for double precision is therefore $t = [(3 \times 8 B + 3 \times 4 B) \times N + (12 \times 8 B + 2 \times 4 B + 14 \times 32 B ) \times M]/(N \times 180 GB/s)$. For the specific problem considered in our benchmarks for this study, only $23\%$ of particles are misplaced, in which case the model should account for $M= 230 000$. We get $t = 0.91 ns$ for double precision and $t = 0.78 ns$ for single precision.
%
%***********************************************************************
\section{Results on GPU}\label{sec:results}
%***********************************************************************
%
In this section, we give performance results using a single GPU with the CRAY OpenACC production compiler. Note that the PGI compiler produces similar results. The benchmark case that will be used for the rest of the paper has a grid of $512 \times 256 \times 1$ with $10^6$ particles, a typical size for realistic simulations. We perform our tests on a CRAY XC30 system having one NVIDIA$^{\circledR}$ Tesla$^{\circledR}$ K20X GPU with ECC enabled (Error-correcting code memory) and running at 745MHz. The host machine uses an Intel$^{\circledR}$ Xeon$^{\circledR}$ E5-2670, running at a clock speed of 2.60GHz. In the following, our measurements are taken with the CRAY OpenACC compiler \texttt{cce/8.3.12} and time reported is per particle per time step. 
%
%-----------------------------------------------------------------
\subsection{Performance of the different charge assignment routines}
%-----------------------------------------------------------------
%
\begin{figure}[hbt]
  \centering 
  \includegraphics[width=0.9\linewidth,keepaspectratio,clip]{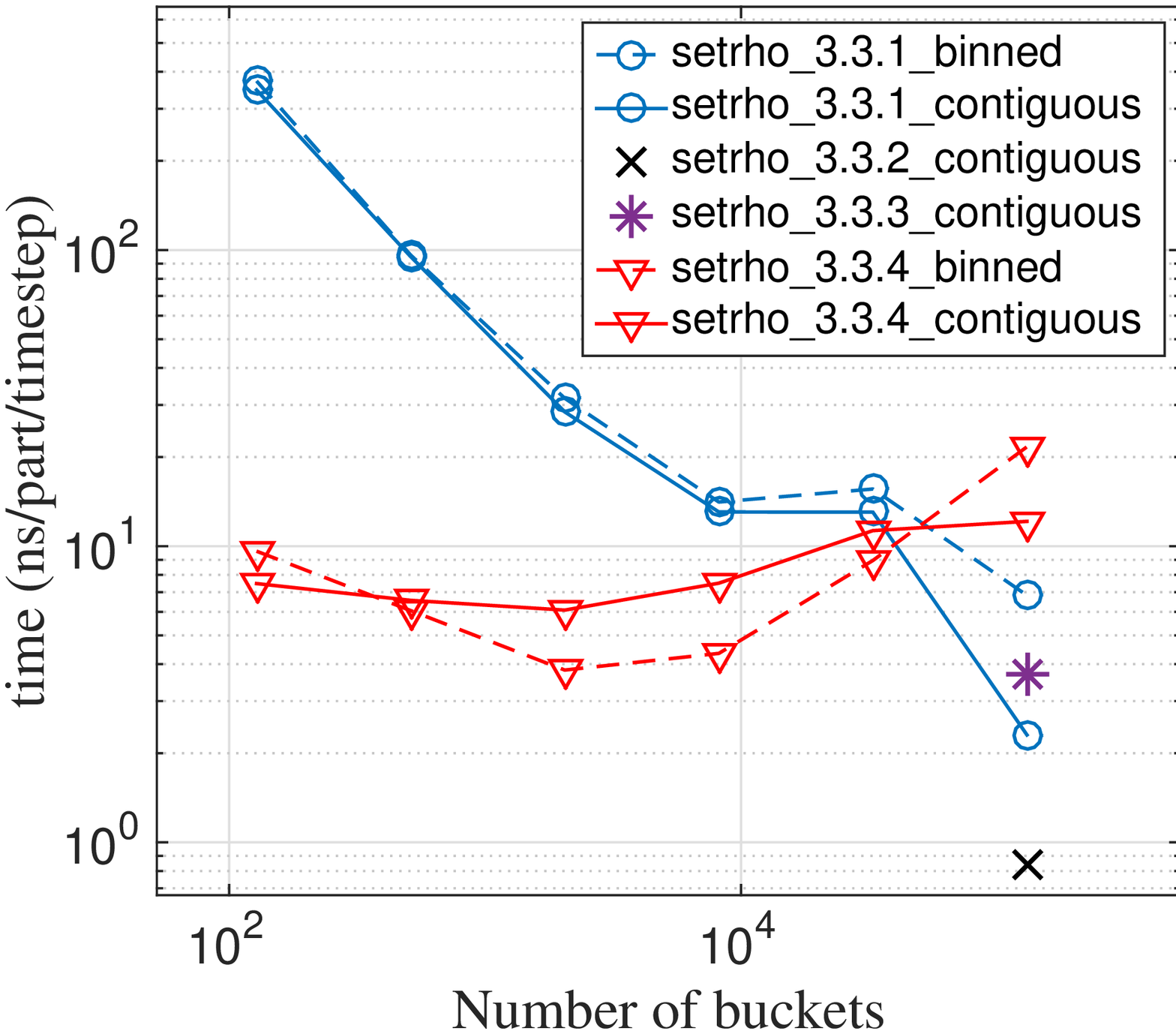}
  \caption{Comparison of the different charge assignment algorithms presented in section~\ref{sec:setrhoroutines} as a function of the total number of buckets}
  \label{fig:setrhoall}
  \end{figure}
The particle-to-grid interpolation step is the most critical for parallelization. We herein study the timings of the different \texttt{setrho} algorithms discussed in sec.~\ref{sec:setrhoroutines}. In this subsection, only double precision will be considered. We plot the total time of this routine in nanoseconds per particle per time step in Fig.~\ref{fig:setrhoall} as a function of the number of buckets. The collision-free algorithm in \texttt{setrho\_3.3.1} with threads on grid points is first considered. We distinguish the contiguous (blue solid line with circles) and the binned (blue dashed line with circles) data structures. As expected, the minimum cost is obtained for the maximum number of bins. For low number of bins, the particle data has to be loaded many times. Each thread will loop over all particles in the bucket, but not all particles contribute to the charge deposition on that particular grid point, except in the case where we sort totally (meaning 1 cell per bucket). This explains the jump that we see in the timing of the blue lines from $256 \times 128$ buckets to $512 \times 256$ buckets. Furthermore, the contiguous version slightly outperforms the binned version in general. It strongly outperforms it for total sorting though (i.e: 1 grid cell/bucket).\\ 

Second, threads on buckets are considered with a contiguous data structure. The collision-resolving \texttt{setrho\_3.3.2} algorithm is represented by a black cross whereas the collision-free algorithm referred to as \texttt{setrho\_3.3.3} is represented by a violet star. Both cases can only be used in the case where particles are fully sorted, that is 1 cell per bucket. On certain physical applications, these algorithms may suffer from load balancing when the buckets are differently filled. Other options for data replication can also be considered.\\ 

Finally, the case where threads are assigned on particles, \texttt{setrho\_3.3.4}, is considered. Here we distinguish the binned data structure (red dashed line with triangles) and the contiguous data structure (red solid line with triangles). The binned version performs at its best for $64 \times 32$ buckets, which is the critical value beyond which the time starts increasing again. There are two possible reasons for this: as the number of particles per bucket gets smaller, any load imbalance between bins becomes more evident, and as the number of bins increases the occupancy of the GPU decreases. The contiguous version does not show better timings except for the maximum number of bins considered when the binned case suffers from load balancing.

\begin{figure}[hbt]
  \centering 
  \includegraphics[width=0.9\linewidth,keepaspectratio,clip]{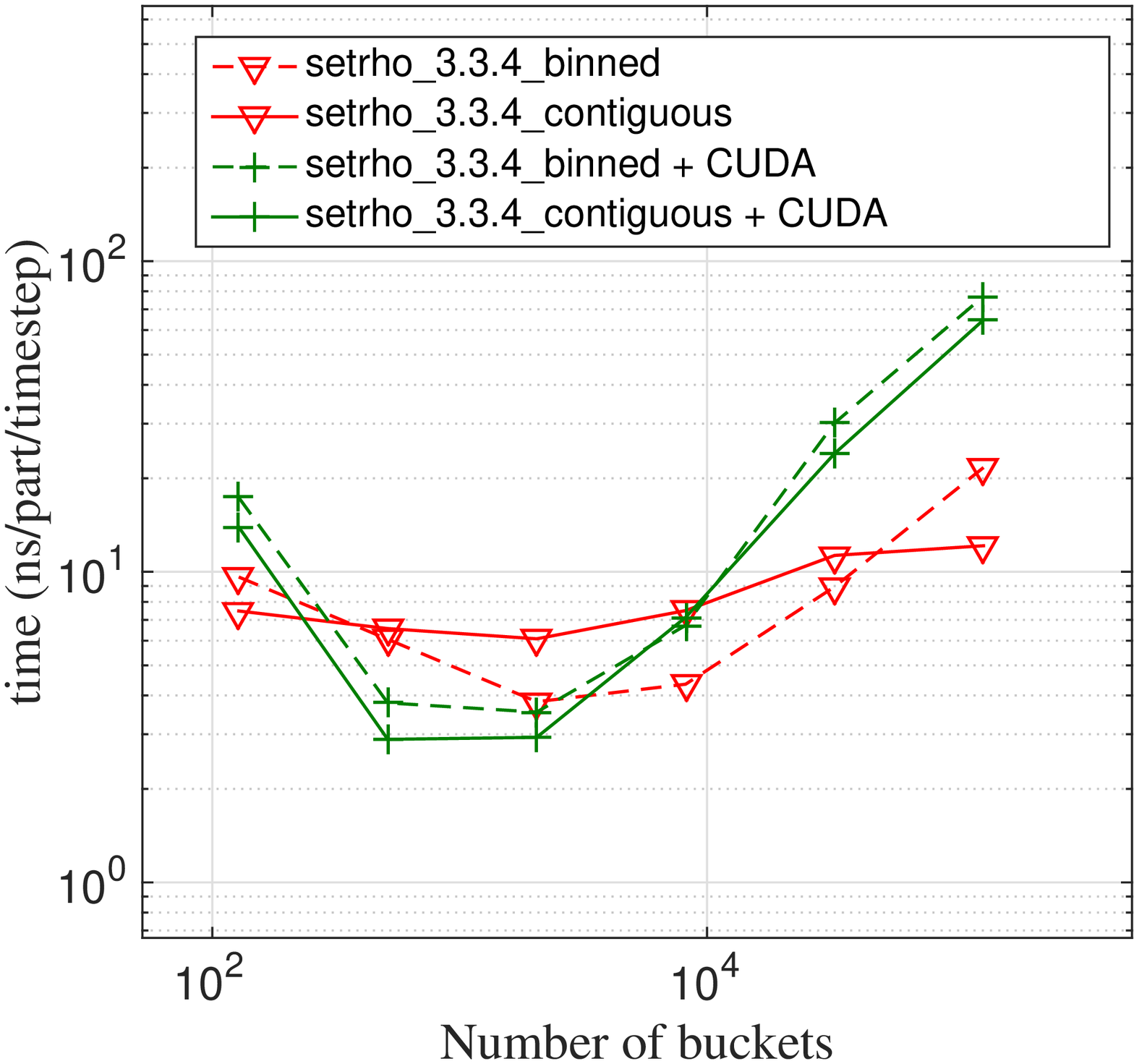}
   \caption{Optimization of the thread on particles charge assignment algorithm using CUDA in the prefixsum and histogram kernels}
  \label{fig:setrhoopti}
  \end{figure}

The comparison among all algorithms for the charge assignment reveals that up to $256 \times 128$ buckets, the threads on grid points algorithm is costly compared to the algorithm having threads on particles. The rightmost points on the graph correspond to the fully sorted case given $512 \times 256$ bins where the best timing is obtained with the collision-resolving threads on bins algorithm \texttt{setrho\_3.3.2}. 
  \begin{figure}[hbt]
  \centering 
  \includegraphics[width=0.9\linewidth,keepaspectratio,clip]{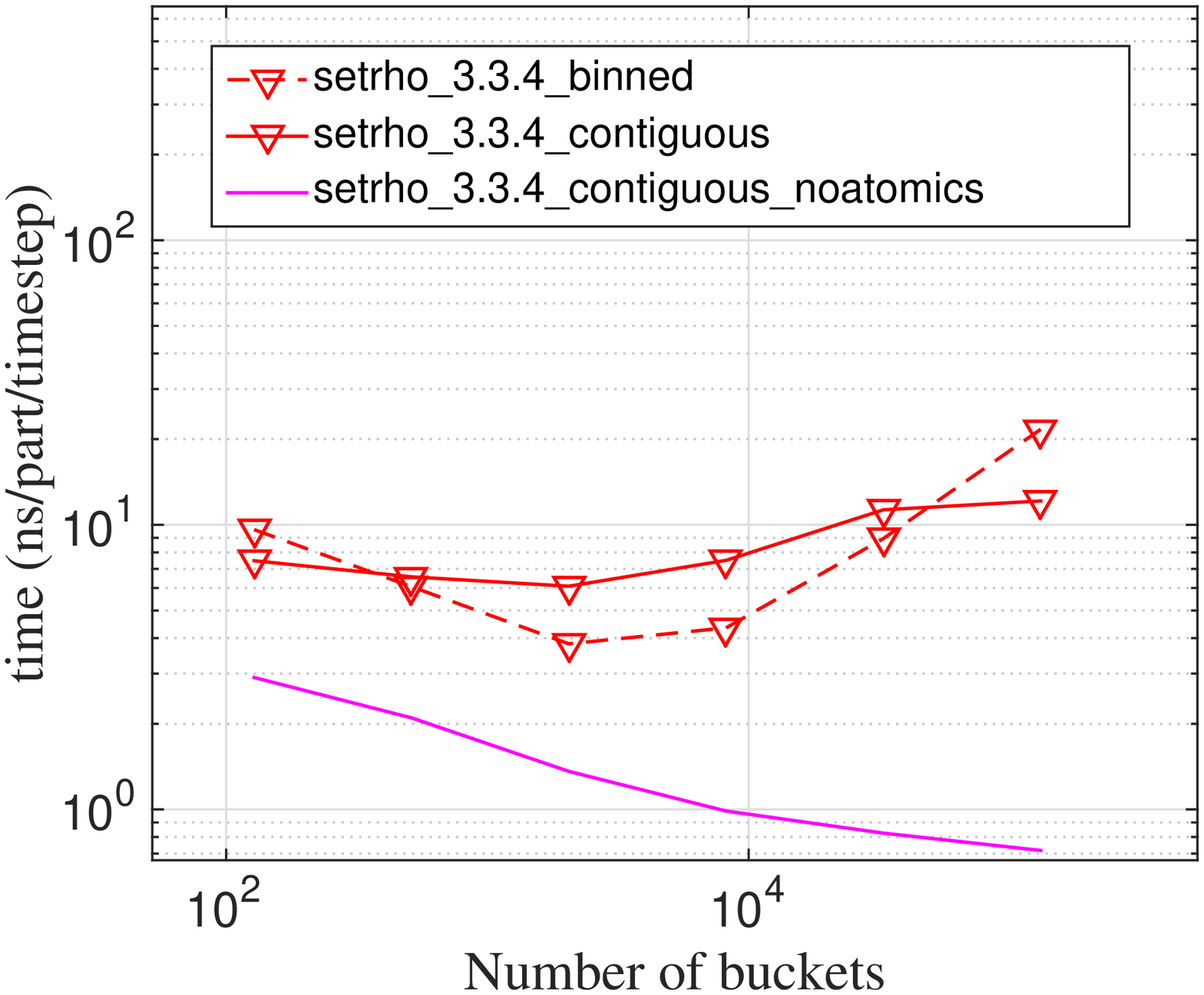}
   \caption{Timings of the charge assignment routine with threads on particles with/without atomic operations revealing the cost of double precision atomics}
  \label{fig:setrhonoatomics}
  \end{figure}
For the rest of the paper, we consider the collision-resolving \texttt{setrho\_3.3.4} algorithm with threads on particles as we believe this option will be the most beneficial one for our ultimate gyrokinetic application. A more detailed argument for this choice will be given in section~\ref{discussion}. The plot in Fig~\ref{fig:setrhonoatomics} shows essentially the binned and contiguous versions of this algorithm (previously shown in Fig.~\ref{fig:setrhoall}) compared to the same contiguous version (purple solid line) where atomics where removed. The cost of atomics increases with the number of buckets since the number of conflicts increases, thus the cost of collisions becomes more evident. An attempt to optimize the contiguous \texttt{setrho} algorithm in CUDA is shown in Fig.~\ref{fig:setrhoopti}. The best timings are obtained with a CUDA optimization for $32 \times 16$ and $64 \times 32$ buckets. This can be explained by the fact that particle data uses shared memory in CUDA. For higher number of buckets, this version becomes more expensive than the pure OpenACC versions. This is mainly due to the cost of the atomics in shared memory.

\FloatBarrier
%
%-----------------------------------------------------------------
\subsection{Performance model compared to numerical experiments}
%-----------------------------------------------------------------
%
 \begin{figure}[h!]
		\centering
        \subfloat[]{\includegraphics[width=0.7\linewidth,keepaspectratio,clip]{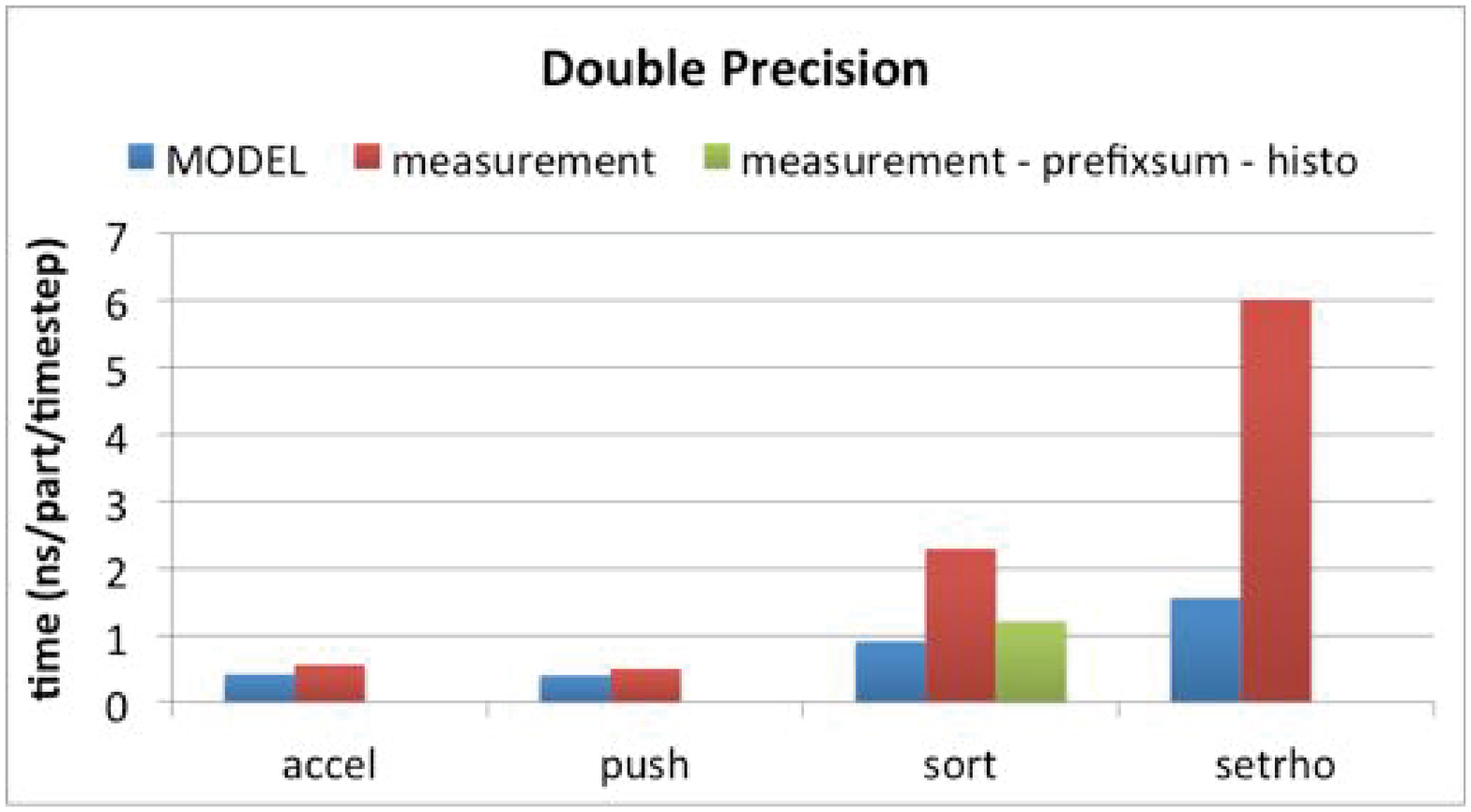}}\\
        \subfloat[]{\includegraphics[width=0.7\linewidth,keepaspectratio,clip]{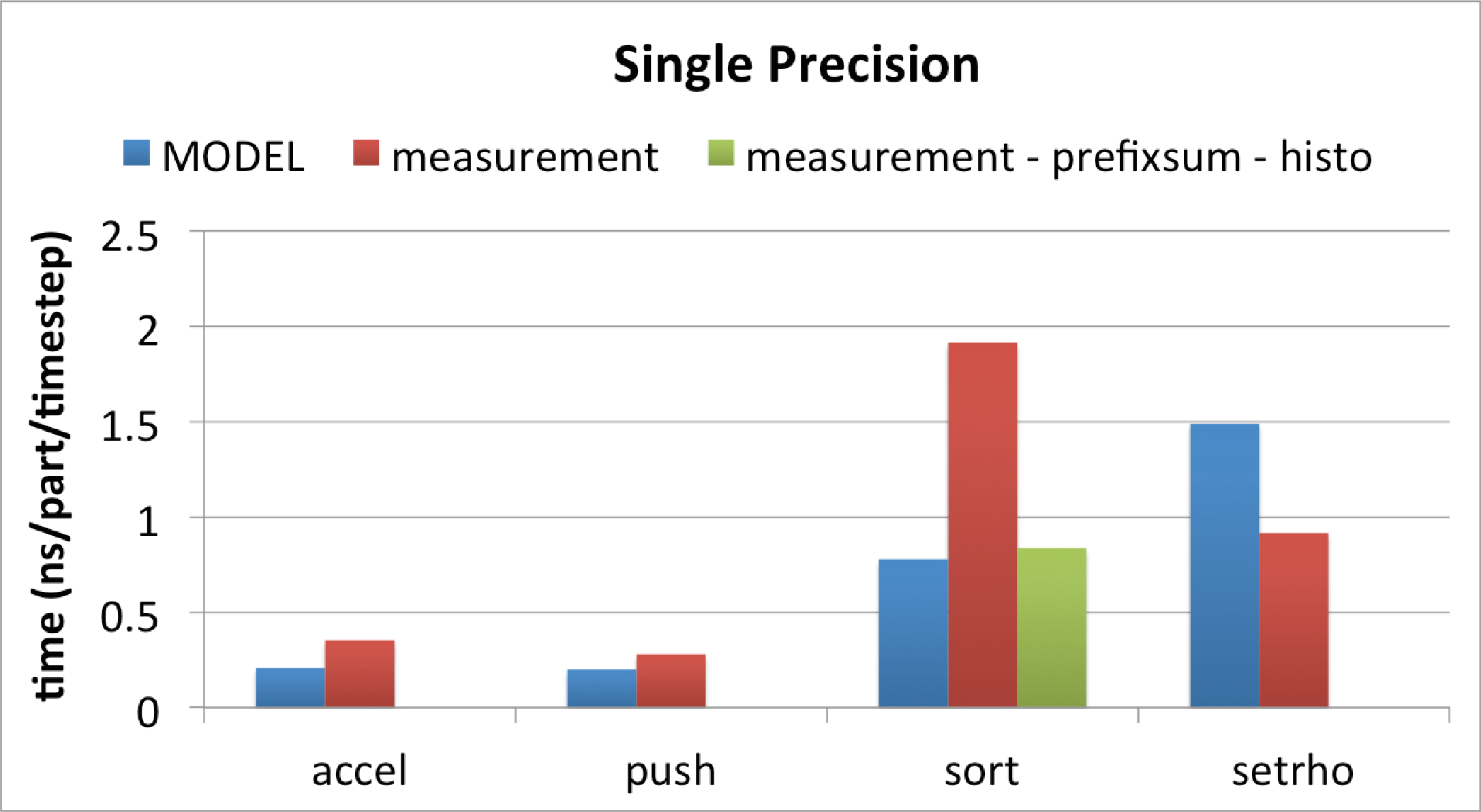}}
        \caption{(a) Performance model in double precision; (b) Performance model in single precision}
        \label{fig:performancemodel}
 \end{figure}
In Fig.~\ref{fig:performancemodel}, we show a comparison between the model and the numerical measurements for the benchmark reference case in both double and single precision. The kernels perform reasonably close to the model in most cases: \texttt{push} and \texttt{accel} are estimated within approximately $60 - 80\%$ in both single and double precision. For \texttt{setrho} we observe that the model overestimates the cost for single precision but underestimates it for double precision. The single precision result is explained by the model not accounting for the impact of caching which makes it more pessimistic than what we can achieve in this case where the particles expose some locality as a result of the sorting. The discrepancy for double precision, on the other hand, is explained by the absence of native double precision atomics on the Kepler architecture. The compiler therefore has to generate a compare and swap loop for the double precision case. Experiments using 64-bit integer atomics as well as the single precision results suggest that a built-in double precision atomic could provide a significant speedup for our application. For the other kernels the model is expected to underestimate the cost since it makes optimistic assumptions (such as 100\% cache hit rate after initial fetch) and only considers the assumed/observed bottleneck. The single precision version of the code, even though used for comparison here, will not be considered for the final benchmarks. This version served for testing the native atomics on the NVIDIA cards.\\

As for the sorting (\texttt{psort}), we recall that prefixsums and histograms are not included in the performance model. Thus, we show in the green bar of Fig.~\ref{fig:performancemodel} the measurement - prefixsum - histogram, which corresponds to the measurement of the total sort minus the measured values for the prefixsums and histograms. The point to make with this is that the model is pretty good for the actually modeled parts of the sorting algorithm, but does not account for all the parts of the algorithm.

As illustrated above, the performance model has various limitations and cannot predict the exact performance figures. Causes are algorithms that are hard to model (such as prefixsums), approximations of the hardware capability and dependence of the performance on the actual data. Still, the benefits are that the model confirms that the observed performance is of the expected order of magnitude. 
%
%-----------------------------------------------------------------
\subsection{Benchmark results: CPU vs GPU}
%-----------------------------------------------------------------
%
 \begin{figure}[h!]
	\centering
       \includegraphics[width=0.8\linewidth,keepaspectratio,clip]{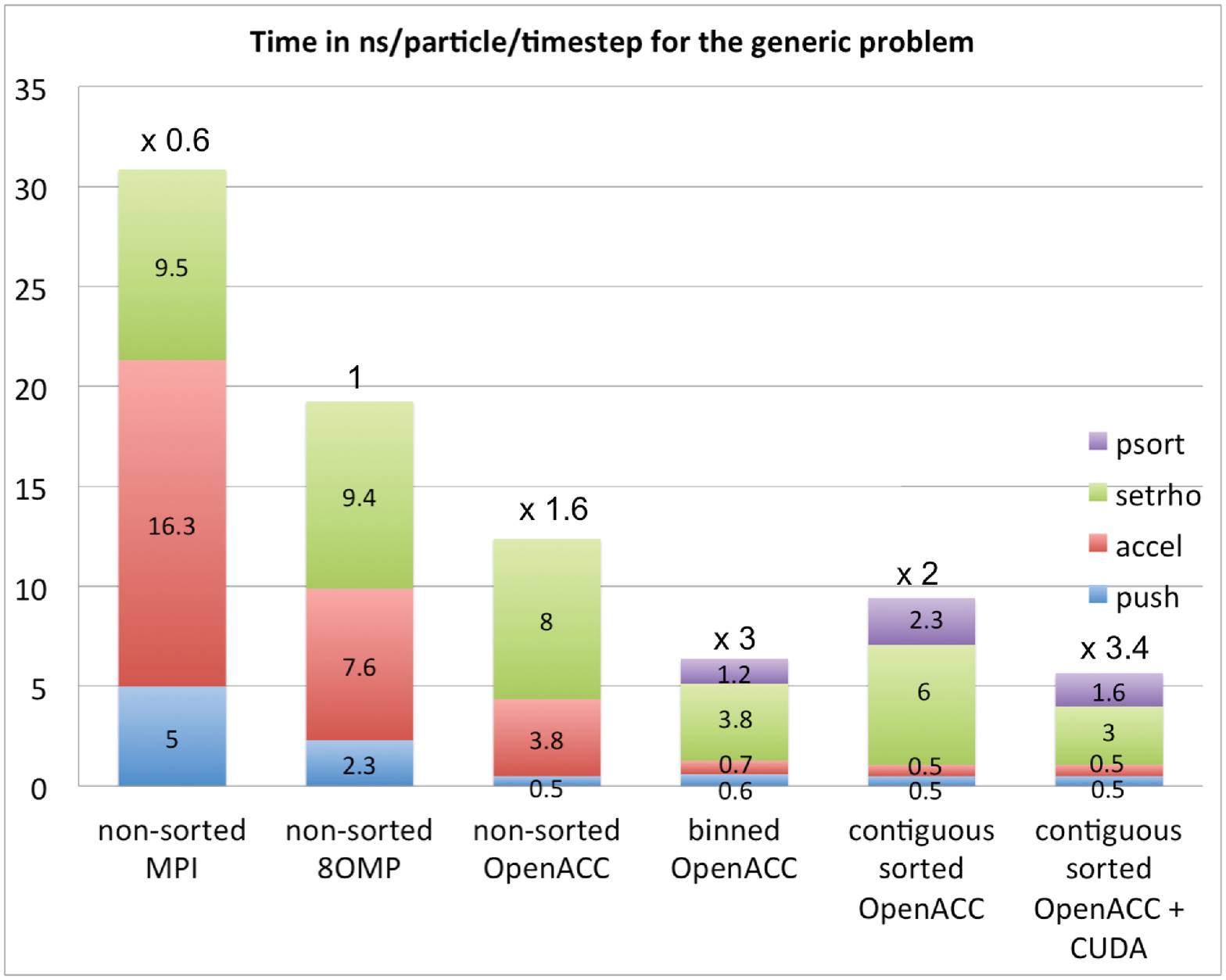}
       \caption{Timings of the \texttt{PIC\_ENGINE}: comparison of various implementations using MPI, OpenMP and OpenACC in double precision (fine tuning optimization in CUDA)}
       \label{fig:generaltimings}
 \end{figure}
We show in Fig.~\ref{fig:generaltimings} the timings of the different implementations using MPI, OpenMP and OpenACC in double precision. CUDA is only used for fine-tuning optimization. The first two columns from the left correspond to the timings of the versions run on CPU only with MPI and OpenMP, respectively, both using 8 cores. It should be noted that these two versions are not sorted on the CPU. We tried applying different sorting algorithms on the CPU for both the MPI and OpenMP implementations. We noted a gain in performance of a factor of $1.7$ for \texttt{setrho} and a factor of $2$ for \texttt{accel}. However, the sorting turned out to be expensive and dominating the total time. Therefore, we observed that there is no gain in total performance on the CPU by sorting the particles. The OpenMP non-sorted implementation will thus be considered as our reference case for the benchmark. Using 8 OpenMP threads, the final time is $19.24 ns$. The MPI implementation is about 2 times slower than the OpenMP reference one. The MPI implementation is based on domain cloning, i.e: field data on the grid (in one case the charge density and the electrostatic field) are replicated and assigned to different CPU cores. Particle data is split among the clones. Charge deposition is made separately in each clone and then grid data has to be reduced across the clones. The OpenMP case does not actually require any replicas of the electric field or charge density.\\
The third column has the timings of the implementation on the GPU using OpenACC without particle sorting. In \texttt{setrho\_3.3.4} we use atomic updates to resolve the conflicts. The gain is mainly noticed in \texttt{push}, not in \texttt{accel} because of the interpolation that requires indirect memory access. In total, we note a $1.6$ times gain in performance with respect to the reference OpenMP CPU version. The fourth column corresponds to timings using binned data structures with OpenACC using the sorting introduced in section~\ref{subsec:sorting} and \texttt{setrho\_3.3.4}. We consider the optimal case with $64 \times 32$ buckets. Even with the cost of the sorting, and the cost of the charge assignment routine being dominant, this implementation on GPU is $3$ times faster than the reference case on a single CPU. Most importantly, if we examine in detail the most relevant routines for this study: \texttt{setrho} is now improved by a factor of $2.4$ and \texttt{accel} is improved by a factor of $10.8$ with respect to the reference version on CPU.\\ 

For the sake of flexibility and simplicity, using the alternative data structure, which consists of keeping the data contiguous and sorting the particles is more advantageous. It has some potential advantages in view of its implementation in application codes. The timings for this implementation are shown in the fifth column. The charge assignment routine dominates the time having a cost double than the one in the binned version. When compared to the reference case, this implementation is still $2$ times faster though. We optimized the prefix sum and the histogram kernels of the charge assignment routine using CUDA with which we managed to reduce the time by a factor of $2$. The main achievement is shown in the 6th column revealing a performance factor $3.4$ times better when executing on the GPU as compared to 8 CPU cores. 
  \begin{figure}[h!]
		\centering
	\includegraphics[width=0.45\linewidth,keepaspectratio,clip]{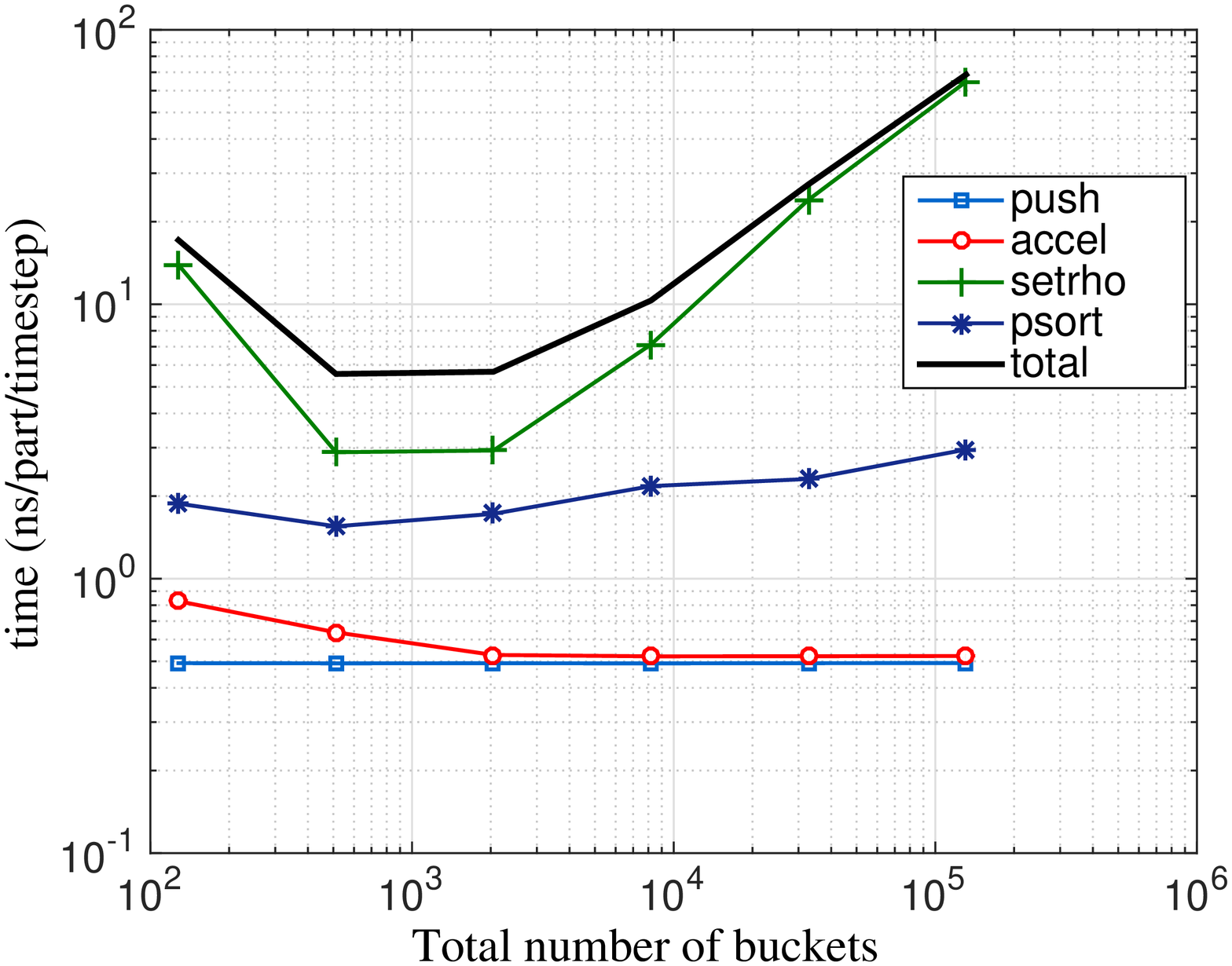}
	\includegraphics[width=0.45\linewidth,keepaspectratio,clip]{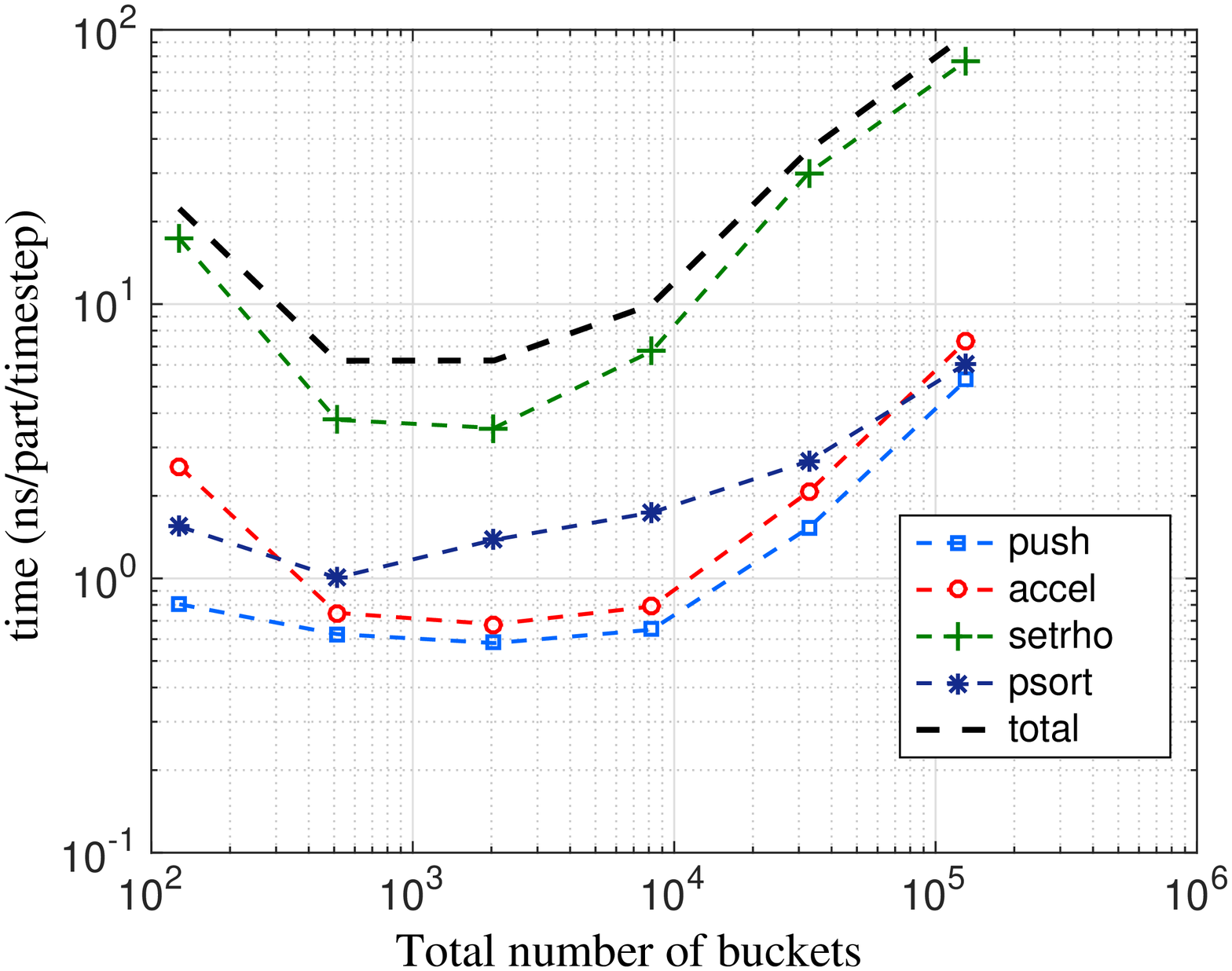}
        \caption{Timings of the \texttt{PIC\_ENGINE} vs number of bins for contiguous (left) and binned (right) data structures}
        \label{fig:alltimings}
 \end{figure}

The timings of the different algorithms in the code using OpenACC and low-level optimization in CUDA for \texttt{setrho\_3.3.4} are plotted in Fig.~\ref{fig:alltimings} as a function of the number of buckets. Both binned and contiguous data structures are considered. Increasing the number of buckets improves data locality, thus leading to better performance for both the contiguous and the binned data structures. For the contiguous case (on the left), we notice that the time of \texttt{push} and \texttt{accel} does not change with increasing number of buckets. Starting from $64 \times 32$ buckets, the cost of \texttt{setrho} increases. This is due to the increasing number of buckets which implies high probability of particle collisions, thus more conflicts. The binned version (on the right) shows a similar total performance with increasing number of buckets and it suffers from load balancing and occupancy starting from $64 \times 32$ buckets. There is an optimum for $64 \times 32$ buckets where we get a comparable time for both contiguous and binned data structures. The sorting algorithm performs at its best for medium grain level. 
%
%***********************************************************************
\section{Discussion in view of physical applications\label{discussion}}
%***********************************************************************
The idea behind this development is to test the most promising algorithms on the basic \texttt{PIC\_ENGINE} in view of an ultimate implementation in various application codes on GPU.  Gyrokinetic codes such as ORB5 code~\cite{tran1998global, jolliet2007global} and GTC~\cite{lin1998turbulent,ethier2005gyrokinetic,ethier2008large} can benefit from lessons learned with this test bed. The final flowchart of the \texttt{PIC\_ENGINE} is shown in~\ref{app_A}.\\
In view of preparing applications requiring many nodes implementation, we have also developed MPI/OpenMP and MPI/OpenACC versions of the code. The \texttt{PIC\_ENGINE} problem, being simple at this stage, is dominated by the time of moving the particles between nodes (\texttt{pmove\_z} routine in the flowchart). This will not be the case in the application code as it will have much more arithmetic operations. The details of our multi-node results will be discussed in further work.\\
 
Of our interest is the optimization of the ORB5 code. Thinking of the steps towards porting it to hybrid machines, we conclude the following from our study.
First, from Fig.~\ref{fig:generaltimings}, we note a factor 2 gain for going from a pure MPI to an OpenMP implementation with 8 threads. The next step for ORB5 will thus be moving to the hybrid MPI/OpenMP implementation. A potential problem that we did not address here is the loss of locality due to gyro-averaging. This will be studied in future work.
Second, ORB5 has many more arithmetic operations especially in \texttt{accel} and \texttt{setrho}, which typically dominate the timings. The promising finding is that \texttt{accel} is improved by a factor of $10.8$ by going from a CPU-only version (8 OpenMP threads) to a GPU version with OpenACC (binned data structure); \texttt{setrho} is also improved by a factor of $2.8$. This means the potential for gain is important in a realistic application, such as ORB5. 

Concerning data structures, the loss of performance in these routines for high number of buckets and binned data structure (Fig.~\ref{fig:alltimings}, right) will become much more important for the overall timings in the application code than for the \texttt{PIC\_ENGINE}. Thus, the contiguous version has probably more chances to suit our application. Moreover, the contiguous data structure is easier to generalize, for instance for higher order elements and more complicated grid structures.\\

Since the performance of charge assignment algorithm, \texttt{setrho}, depends on the problem size and the architecture, we would like to test all the possible algorithms in our application. According to section~\ref{sec:setrhoroutines} and Fig.~\ref{fig:setrhoall}, the most promising schemes may be (i) threads on particles with partial bucket sorting (algorithm~\ref{subsub:threadsparticles}); (ii) threads on grids with full sort of particles and contiguous data (algorithm~\ref{subsub:threadsgrid}); (iii) collision-resolving threads on bins (algorithm~\ref{subsub:threadsbins}), but it is not sure that the shared memory size will be sufficient for higher order interpolation. Applications that use linear order interpolation can benefit from this last option. 
%
%***********************************************************************
\section{Conclusions\label{conclusions}}
%***********************************************************************
%
A hard problem in modern computing is how to deal with legacy codes so that they can benefit from the current and future High Performance Computing (HPC) resources. The challenge became apparent with the advent of hybrid architectures offering massive amounts of multithreading. An integrated programming environment, preferably applicable to both traditional multicore HPC systems as well as to new hybrid CPU/GPU architectures, is therefore key in the quest for
exascale computing.\\ 

The main objective behind our work is the development of a platform, that we refer to as a \texttt{PIC\_ENGINE}, for testing the performance of PIC algorithms on hybrid CPU+GPU and many-core platforms CPU+ Xeon Phi. This permits a valuable testing ground for the algorithmic choices in a more efficient way than on a full application production code. Of course, depending on the physical application, different algorithms are best suited for different hardware. This paper focuses on CPU/GPU systems. The \texttt{PIC\_ENGINE} can be run on the Intel Many Integrated Core architecture (Xeon Phi) as well. Results on the latter architecture are subject to ongoing efforts that will be shown in a separate work.\\

In this paper, we considered a simple Vlasov electrostatic problem in 6D phase space using Cartesian coordinates and linear particle-to-grid interpolations. For the sake of simplicity, the field-solver was not included and particles move into a frozen given electric field. We took the option to rely on the OpenACC programming standard which allows existing HPC applications to be ported to run on one or more GPUs with only modest changes to the original code. It also offers more portability and maintainability for the code base. Most importantly, since the tools for debugging on the CPU are currently better supported, the same code can still be run efficiently on a pure CPU-based system. We focused on single node performance and compared CPU and GPU implementations. An extension to multiple GPUs on a drift-kinetic problem is currently being carried out and will be evaluated in future work.\\

We discussed several  algorithms for particle sorting and charge deposition using two main data structures: binned and contiguous. Two ways of storing data were considered: Array of structures (AOS) and Structure of arrays (SOA). We found that SOA gives higher performance on both the CPU and the GPU.\\

Different options were considered for the charge deposition algorithm \texttt{setrho}. Associating the threads to the grid is a choice where no race condition can happen: this is a collision-free algorithm. Associating threads with bins of particles and grid data replication (local grids) is a second collision-free option. In this case, one can either use atomics for the inner particle loop, or use additional local temporary variables without atomics. The last option is to assign threads with particles and resolve collisions with atomic operations. In this latter option, a hand-optimized low-level version with CUDA allows using the GPU shared memory as an intermediate step in the reduction.\\

Various sorting algorithms were also implemented and tested on both the CPU and the GPU. Sorting did not improve the overall performance on the CPU, but it did so on the GPU. The sorting algorithm proposed in~\cite{jocksch2015bucket} and described in section~\ref{subsec:sorting} requires a smaller memory footprint and shows a better performance for cases where a fraction of particles less than $50\%$ have to be moved at every timestep. There is an optimum number of buckets where this sorting is beneficial.\\ 

For the given benchmark platform, we have a theoretical understanding of the measurements showing a significant impact of non-nativeness of atomics for double precision data. Results show that the code performs reasonably well compared to this idealized model, i.e: roughly $80\%$ of the performance is achieved except for the charge assignment routine. The latter does not perform as expected by the model mainly due to the lack of native atomic operations in double precision. In single precision, we show a much improved performance in \texttt{setrho} as the NVIDIA card supports single precision atomics.\\

Finally, our numerical experiments on a single node show that a performance gain of a factor of $3$ can be obtained by porting the PIC algorithm on the GPU using only OpenACC. The algorithms we presented are scalable with size of the mesh and can be easily ported to execute on multiple GPUs. The use of shared memory is, however, critical for optimal performance. CUDA is therefore employed for low-level optimization in such a way to use shared memory with atomics. Our results show that the optimised GPU version of the code is $3.4$ times faster than on 8-core CPU implementation. 
%
%***********************************************************************
\section*{Acknowledgments}
%***********************************************************************
The authors would like to thank O. Sauter for helpful discussions. This project has been supported by the Platform for Advanced Computing (PASC) programme. It is within the framework of the EUROfusion Consortium and has received funding from the Euratom research and training programme 2014-2018 under grant agreement No 633053. The views and opinions expressed herein do not necessarily reflect those of the European Commission.

%***********************************************************************
% Appendices 
%***********************************************************************
\newpage
%%%%%
\appendix
\section{Flowchart of the \texttt{PIC\_ENGINE}\label{app_A}}
\begin{figure}[hbt]
  \centering
  \includegraphics[width=0.53\linewidth,keepaspectratio,clip]{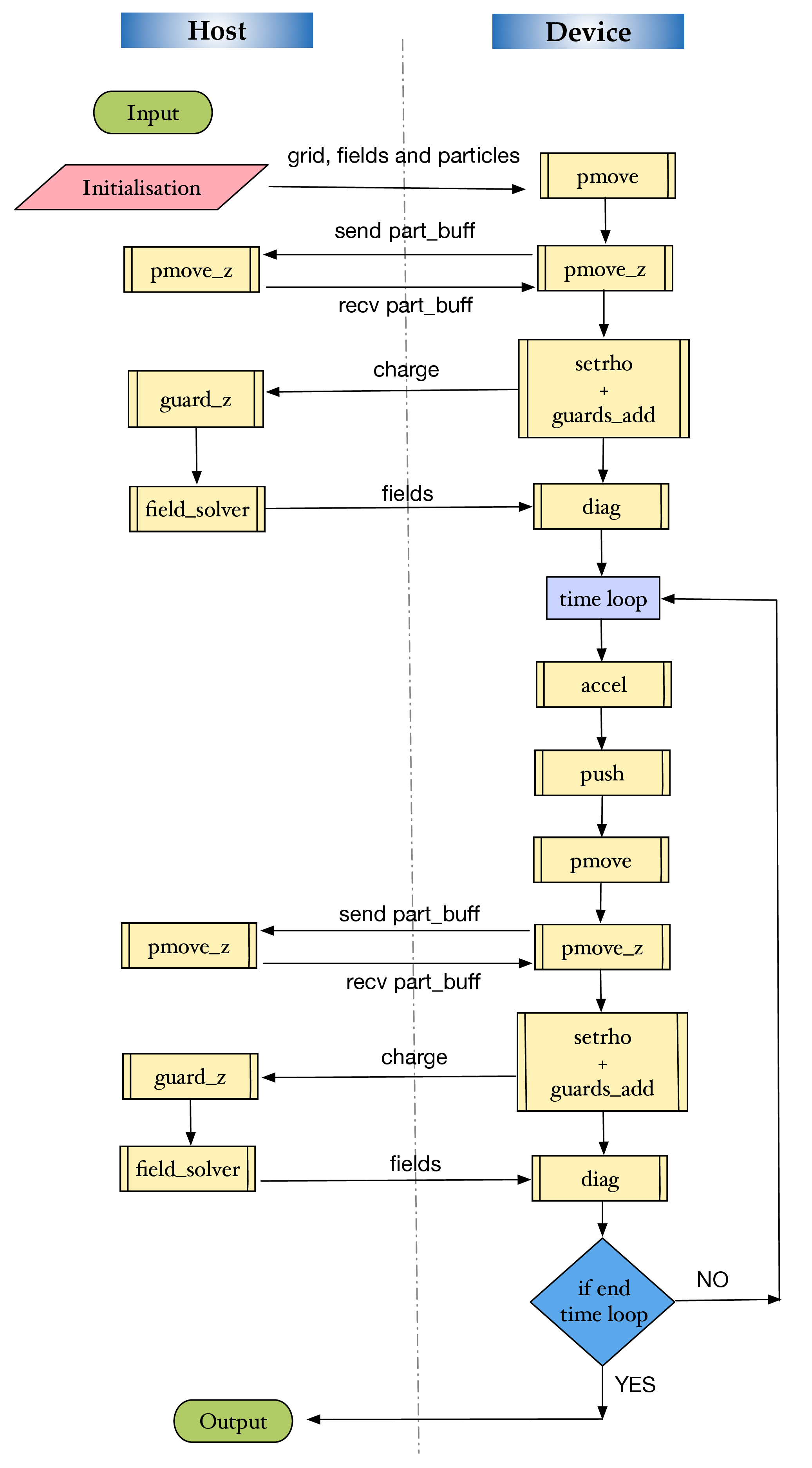}
  \caption{Flowchart of the \texttt{PIC\_ENGINE} showing the main subroutines.}
  \label{fig:flowchart}
\end{figure}
%%%%%%
\newpage
%%%%%%
\section{Fortran examples for accelerating the charge deposition routine\label{app_B}}
An example of the \texttt{setrho()} routine with parallelization over grid points (subsection~\ref{subsub:threadsgrid}) while considering binned particle data.\\

\begin{verbatim}
!$acc parallel present(part_att,xgrid,ygrid,zgrid,rho_glob)
!$acc loop gang worker vector private(ip,ix,iy,iz,ibin,wx,wy,wz), collapse(3)
do k=0,nz
  do j=0,ny
    do i=0,nx
      do over bins adjacent to grid point j
        do over bins adjacent to grid point  i
          do ip = 1,npbin(bin)
              wx = part_att(ip,1,ibin)-xgrid(ix)
              wy = part_att(ip,2,ibin)-ygrid(iy)
              wz = part_att(ip,3,ibin)-zgrid(iz)
              IF(ABS(wx).LE.1.d0 .AND. ABS(wy).LE.1.d0 .AND. ABS(wz).LE.1.d0) THEN
!$acc atomic
                 rho_glob(ix,iy,iz) = rho_glob(ix,iy,iz) + &
                  & MIN(1.d0-wx,1.d0+wx)*MIN(1.d0-wy,1.d0+wy)*MIN(1.d0-wz,1.d0+wz)
               END IF
          end do
        end do
      end do
    end do
  end do
end do
!$acc end parallel loop
\end{verbatim}
To offload a region of the code to the device, one should add directives to the source code. For performance reasons, the main program runs on the host and compute-intensive regions are offloaded to the GPU. The execution and data management is guided by the programmer using OpenACC directives. In the example above, we used the \texttt{parallel} directive which indicates that the following region of code can be accelerated. Then we specify through the \texttt{present} clause which variables are already present on the accelerator. A \texttt{loop} construct is applied to the immediately following nested loops. The loop iterations will then be divided over the threads of the GPU by the compiler. These threads are organized in a three-level hierarchy in the GPU hardware. The so-called \texttt{gangs} are at the highest level. Within these are \texttt{workers} made up of \texttt{vectors}. Therefore, \texttt{gang} parallelism is coarse-grain, \texttt{worker} parallelism is fine-grain where each gang has one or more workers. Within the parallel region, a \texttt{loop} directive instructs the work sharing among the accelerator's workers. We insert additional clauses in the \texttt{loop} directive to optimize the implicit data management chosen by the compiler.  Here we used the three levels, \texttt{gang worker vector}, then we specified as \texttt{private} the variables that need to be created as private copies for each thread that executes the loop. The \texttt{collapse(3)} clause applies the \texttt{loop} directive to the 3 tightly nested loops. Finally,  \texttt{acc atomic} is used in order to resolve the memory collisions in updating the charge density. 

It is important to note that calls to CUDA kernels and libraries can be made within regions accessible by the accelerator. These may be used to hand-optimize low-level versions of performance-critical routines (such as the charge deposition routine). In a CUDA code, the \texttt{setrho()} routine is more complex and is harder to understand. CUDA has been used to optimize the \texttt{setrho()} routine (subsection~\ref{subsub:threadsparticles}) which we show hereafter:

\begin{verbatim}
global__ void setrhok_bin(int tot_bins, int mnpb, int *npbin, real *part_att, int nx, 
real *xgrid, int ny, real *ygrid, int nz, real *zgrid, real *rho, int nsplitx, int nsplity){
extern __shared__ real rholoc[];
real wx, wy, wz;
int ix, iy, iz, ixl, iyl, ixb, iyb, i, ixl2, iyl2, iz2, nlx, nly, bin, ip, ibin, jbin;
bin=blockIdx.x/((mnpb-1)/blockDim.x+1);
if (npbin[bin]>(blockIdx.x%((mnpb-1)/blockDim.x+1))*blockDim.x){
 ip=(blockIdx.x%((mnpb-1)/blockDim.x+1))*blockDim.x+threadIdx.x;
 nlx=nx/nsplitx;
 nly=ny/nsplity;
 ibin = bin%nsplitx;
 jbin = bin/nsplitx;
  if (ip<npbin[bin]){
    ix = part_att[bin*6*mnpb+0*mnpb+ip];
  if (ix<0) ix=0; if (ix>nx-1) ix=nx-1;
     iy = part_att[bin*6*mnpb+1*mnpb+ip];
  if (iy<0) iy=0; if (iy>ny-1) iy=ny-1;
     iz = part_att[bin*6*mnpb+2*mnpb+ip];
  if (iz<0) iz=0; if (iz>nz-1) iz=nz-1;
     wx = part_att[bin*6*mnpb+0*mnpb+ip]-xgrid[ix];
     wy = part_att[bin*6*mnpb+1*mnpb+ip]-ygrid[iy];
     wz = part_att[bin*6*mnpb+2*mnpb+ip]-zgrid[iz];
     ixl  =(ix+nlx)%nlx; iyl=(iy+nly)%nly;
     ixb =ibin; iyb=jbin;
     }
for (i=threadIdx.x; i<(nlx+1)*(nly+1)*(nz+1); i+=blockDim.x){
        rholoc[i]=0.;
     }
    syncthreads();
 if (ip<npbin[bin]){
    atomicAdd(&rholoc[ixl+(nlx+1)*iyl+(nlx+1)*(nly+1)*iz], (1.0-wx)*(1.0-wy)*(1.0-wz));
    atomicAdd(&rholoc[(ixl+1)+(nlx+1)*iyl+(nlx+1)*(nly+1)*iz], wx*(1.0-wy)*(1.0-wz));
    atomicAdd(&rholoc[ixl+(nlx+1)*(iyl+1)+(nlx+1)*(nly+1)*iz], (1.0-wx)*wy*(1.0-wz));
    atomicAdd(&rholoc[ixl+(nlx+1)*iyl+(nlx+1)*(nly+1)*(iz+1)], (1.0-wx)*(1.0-wy)*wz);
    atomicAdd(&rholoc[(ixl+1)+(nlx+1)*(iyl+1)+(nlx+1)*(nly+1)*iz], wx*wy*(1.0-wz));
    atomicAdd(&rholoc[(ixl+1)+(nlx+1)*iyl+(nlx+1)*(nly+1)*(iz+1)], wx*(1.0-wy)*wz);
    atomicAdd(&rholoc[ixl+(nlx+1)*(iyl+1)+(nlx+1)*(nly+1)*(iz+1)], (1.0-wx)*wy*wz);
    atomicAdd(&rholoc[(ixl+1)+(nlx+1)*(iyl+1)+(nlx+1)*(nly+1)*(iz+1)], wx*wy*wz);
     }
    syncthreads();
for (i=threadIdx.x; i<(nlx+1)*(nly+1)*(nz+1); i+=blockDim.x){
   ixl2=i%(nlx+1); iyl2=(i/(nlx+1))%(nly+1); iz2=i/(nlx+1)/(nly+1);
   atomicAdd(&rho[(ixb*nlx+ixl2)+nx*(iyb*nly+iyl2)+nx*ny*iz2], rholoc[i]);
     }
  }
}
\end{verbatim}

\FloatBarrier
%
%***********************************************************************
%% References
%***********************************************************************

%% Following citation commands can be used in the body text:
%% Usage of \cite is as follows:
%%   \cite{key}         ==>>  [#]
%%   \cite[chap. 2]{key} ==>> [#, chap. 2]
%% References with bibTeX database:
\bibliographystyle{elsarticle-num} % or unsrt
\bibliography{bibmain}

\end{document}